\documentclass[aps,prd,showpacs,superscriptaddress,nofootinbib,floatfix,10pt]{revtex4-2} 
\usepackage[utf8]{inputenc}
\usepackage{color}
\usepackage{amsfonts,amsmath,amssymb,leftindex}

\usepackage{longtable,booktabs}
\usepackage{graphicx}
\usepackage{bm}
\usepackage[colorlinks=true,linkcolor=blue,citecolor=blue]{hyperref}
\usepackage{geometry}
\usepackage{url}
\usepackage{epstopdf}
\usepackage{mathtools}
\usepackage{subfigure}
\usepackage{multirow}
\usepackage{diagbox}
\usepackage{afterpage}
\usepackage{placeins}
\usepackage{float}
\usepackage{mathrsfs}
\usepackage{setspace}
\usepackage{orcidlink}
\allowdisplaybreaks[1]
\graphicspath{{figs/}}

\renewcommand{\arraystretch}{1.5}

\newcommand{\zjwl}{\affiliation{College of Information and Intelligence Engineering, Zhejiang Wanli University, Zhejiang 315101, China}}

\newcommand{\nbu}{\affiliation{Physics Department, Ningbo University, Zhejiang 315211, China}}

\newcommand{\bnu}{\affiliation{School of Physics and astronomy, Beijing Normal University, Beijing 100875, China}}

\newcommand{\hnue}{\affiliation{College of Science, Henan University of Engineering, 451191 Zhengzhou, China}}

\newcommand{\usc}{\affiliation{School of Nuclear Science and Technology, University of South China, Hengyang, 421001, Hunan, China}}
\begin{document}

\title{Probing the soft rescattering parameters in $B$ decays involving a scalar meson with QCD factorization}

\author{Jing-Juan Qi} 
\email{jjqi@mail.bnu.edu.cn}
\zjwl\nbu

\author{Zhen-Yang Wang} 
\email{wangzhenyang@nbu.edu.cn}
\nbu

\author{Zhen-Hua Zhang} 
\email{zhangzh@usc.edu.cn}
\usc

\author{Ke-Wei Wei} 
\email{weikw@hotmail.com}
\hnue

\author{Xin-Heng Guo} 
\email{xhguo@bnu.edu.cn}
\bnu
\date{\today\\}
\begin{abstract}
In this work, the soft rescattering parameters in the $B^\pm\rightarrow \pi^\pm\pi^+\pi^-$ and $B^\pm\rightarrow K^\pm\pi^+\pi^-$ decays with the light scalar meson $f_0(500)$ as the intermediate resonance are studied within the QCD factorization. Considering the interference effect between $\rho(770)^0$ and $f_0(500)$, we utilize the experimentally more direct event yields for fitting and get the soft rescattering parameters $|\rho_k^{SP}|=3.29\pm1.01$ and $|\rho_k^{PS}|=2.33\pm0.73$ in $B\rightarrow PS$ and $B\rightarrow SP$ decays ($P$ and $S$ denote pseudoscalar and scalar mesons, respectively), respectively. We also study the branching ratios and $CP$ asymmetries in the decay modes involving other scalar mesons, including $f_0(980)$, $a_0(980)$, $a_0(1450)$ and $K_0^*(1430)$, to test the rationality of the values of $|\rho_k^{SP}|$ and $|\rho_k^{PS}|$. Meanwhile, the wealth of experimental data facilitate the examination of the forward-backward asymmetry induced $CP$ asymmetries (FB-CPAs), and the localized $CP$ asymmetries (LACPs). We investigate these asymmetries resulting from the interference between $\rho(770)^0$ and $f_0(500)$ for $B^\pm\rightarrow \pi^\pm\pi^+\pi^-$ and $B^\pm\rightarrow K^\pm\pi^+\pi^-$ decays when  the invariant mass of $\pi^+\pi^-$ locates in the low-energy region $0.445\mathrm{GeV}<m_{\pi\pi}<0.795\mathrm{GeV}$. Our theoretical results of FB-CPAs and LACPs align with the experimental findings. We propose that the interference between $\rho(770)^0$ and $f_0(500)$ can be extended to other beauty and charmed mesons decays.
\end{abstract}

\maketitle
\newpage

\section{Introduction}
The inclusive $CP$ asymmetries of $B^\pm\rightarrow \pi^\pm \pi^+\pi^-$, $B^\pm\rightarrow K^\pm \pi^+\pi^-$ and some other charmless three-body decays were measured in 2014 \cite{LHCb:2014mir}. The dynamical origin of these $CP$ asymmetries can be better understood with the amplitude analyse of decay channels. These analyse were based on a data sample corresponding to an integrated luminosity of  3 f$b^{-1}$ of $pp$ collisions recorded with the LHCb detector. The distributions of $CP$ asymmetries in the Dalitz plots were also studied and the LHCb Collaboration proposed two different sources of $CP$ violation.  The first one may be associated with the $\pi^+\pi^-\rightarrow K^+K^-$ rescattering strong-phase difference in the region around 1.0 to 1.5 $\mathrm{GeV}$ \cite{LHCb:2013ptu,LHCb:2013lcl}, while the second one, observed in both $B^\pm\rightarrow \pi^\pm \pi^+\pi^-$  and $B^\pm\rightarrow K^\pm \pi^+\pi^-$ decays around the $\rho(770)^0$ mass region, can be attributed to the final-state interference between the $S$-wave and $P$-wave components in the Dalitz plots. In 2020, the LHCb Collaboration conducted a more detailed amplitude analysis of the three-body decay $B^\pm\rightarrow \pi^\pm\pi^+\pi^-$ and reported their new findings \cite{LHCb:2019sus,LHCb:2019jta}.  They used three different approaches (the isobar model, the $K$-matrix model, and a quasi-model-independent binned approach) to study the $S$-wave component of this decay. According to the isobar model, the $S$-wave fit fraction was approximately $0.25$ which is mainly influenced by the $f_0(500)$ resonance and  significant $CP$ violation was observed due to the interference between the $P$-wave and $S$-wave components. Based on the theoretical analysis, the aforementioned important $CP$ violation is attributed to the interference between $\rho(770)^0$ and $f_0(500)$, which  can well be explained based on a QCD factorization approach for weak decays of $B$ meson.

The decays of $B$ mesons offer a rich testing ground for exploring the flavor pictures of the Standard Model and probing the  potential hints of new physics. To ensure reliable predictions, a key aspect involves evaluating the short-distance QCD corrections to the hadronic matrix elements of $B$ decays. Various theoretical approaches, such as QCD factorization (QCDF) \cite{Beneke:1999br,Beneke:2003zv}, perturbative QCD (pQCD)\cite{Keum:2000ph,Keum:2000wi}, and the soft-collinear effective theory (SCET) \cite{Bauer:2000ew,Bauer:2001yt}, are extensively employed  to calculate the decay amplitudes of $B$ mesons. In the QCDF, the calculation of hard spectator and weak annihilation corrections always suffers from the divergence at the end-point of convolution integrals of meson's light-cone distribution amplitudes. To estimate the end-point contribution, phenomenological parameter $X_k$ is introduced, which includes the strength parameter $\rho_k$ and the possible strong phase parameter $\phi_k$. Both of them are utterly unknown from the first principles of QCD dynamics, and thus can only be obtained through fitting the experimental data. It is precisely because of these two soft rescattering parameters, the hard spectator and weak annihilation contributions also play an indispensable role for evaluating the $CP$ asymmetry.

Recently, the new observables for the the forward-backward asymmetries (FBAs) and forward-backward asymmetry induced $CP$ asymmetries (FBA-CPAs) were proposed \cite{Wei:2022zuf,Hu:2022eql} for the $B$ meson decays. In fact,  FBAs have been recognized as valuable observables since as early as 1956, for example, in the discovery of the parity violation of weak interaction \cite{Lee:1956qn,Wu:1957my}, the precision measurement of the $Z$ boson \cite{ALEPH:2005ab}, and the study of the lepton universality \cite{Ali:1991is,Belle:2009zue}. However, they relatively rarely appear in studies of beauty and charmed meson decays. The FBA-CPA can be used to  isolate $CP$ violation caused by the interference between adjacent resonance states and can be directly obtained based on the reported number of events, demonstrating its apparent advantages of directness and efficiency. Since the FBA-CPA has not been studied widely for non-leptonic decays of $B$ and $D$ mesons, it is urgent to investigate this important observable in large experiments like LHCb.

This paper is structured as follows. In the next section, we review the formalism for $B$ decays in the QCD factorization approach. In Sect. ${\mathrm{\uppercase\expandafter{\romannumeral3}}}$, we present the definition of the FBA and FBA-CPA for three-body decays of the $B$ meson. The numerical results are given in Sect. ${\mathrm{\uppercase\expandafter{\romannumeral4}}}$ and we summarize and discuss our work in Sect ${\mathrm{\uppercase\expandafter{\romannumeral5}}}$. In Appendix,  we give the amplitudes of the $B$ meson decays and the propagators of the intermediate resonances.

\section{B DECAYS IN THE QCD FACTORIZATION}
In the following, we provide a brief overview of the non-annihilation and annihilation amplitudes for $B$ two-body decays utilizing the QCDF approach. For more detailed information, please refer to Refs. \cite{Beneke:2003zv,Beneke:2001ev}. The effective weak Hamiltonian for charmless hadronic $B$ decays can be represented as
\begin{equation}\label{Hamiltonian}
\mathcal{H}_{eff}=\frac{G_F}{\sqrt{2}}\sum_{i=1}^{10}\sum_{p=u,c}\lambda_p^{(D)}C_i(\mu)Q^q_i(\mu),
\end{equation}
where $G_F$ is the Fermi constant, $\lambda_p^{(D)}=V_{pb}^*V_{pD}$ ($D=d$ or $s$) is the Cabibbo-Kobayashi-Maskawa (CKM) factor, and $C_i(\mu)$ is the Wilson coefficient which is perturbatively calculable from first principles, the four-quark effective operators $Q_{1,2}^q$, $Q_{3,\cdot\cdot\cdot,6}$ and $Q_{7,\cdot\cdot\cdot,10}$ are tree-level, QCD penguin, and electroweak penguin operators, respectively. The non-annihilation and annihilation amplitudes can be influenced by these effective operators.
The effective Hamiltonian matrix elements can be
written as
\begin{equation}\label{Amplitude}
\langle M_1M_2|\mathcal{H}_{eff}|B\rangle=\sum_{p=u,c}\lambda_p^{(D)}\langle M_1M_2|\mathcal{T}_A^p+\mathcal{T}_B^p|B\rangle,
\end{equation}
where $\mathcal{T}_A^p$ and $\mathcal{T}_B^p$ describe the contributions from non-annihilation and annihilation amplitudes, respectively, which can
be expressed in terms of $a_i^p$ and $b_i^p$. Generally, $a_i^p$ includes the contributions from naive factorization, vertex correction, penguin amplitude and
spectator scattering, and can be expressed as follows \cite{Beneke:2003zv}
\begin{equation}\label{a}
\begin{split}
a_i^p{(M_1M_2)}&={(C_i+\frac{C_{i\pm1}}{N_c})}N_i{(M_2)}+\frac{C_{i\pm1}}{N_c}\frac{C_F\alpha_s}{4\pi}{\bigg[V_i{(M_2)}+\frac{4\pi^2}{N_c}H_i{(M_1M_2)}\bigg]+P_i^p{(M_2)}},
\end{split}
\end{equation}
where $C_i$ are the Wilson coefficients, the upper (lower) signs apply when $i$ is odd (even), $N_i{(M_2)}$ are leading-order coefficients, $C_F={(N_c^2-1)}/{2N_c}$ with $N_c=3$, the quantities $V_i{(M_2)}$ account for one-loop vertex corrections, $H_i{(M_1M_2)}$ describe hard spectator interactions with a hard gluon exchange between the emitted meson and the spectator quark of the $B$ meson, and $P_i^p{(M_1M_2)}$ are from penguin contractions \cite{Beneke:2003zv}.
When $M_1M_2=VP,PV$ ($V$ and $P$ denote vector and pseudoscalar mesons, respectively.), the correction from the hard gluon exchange between $M_2$ and the spectator quark is given by \cite{Beneke:2003zv,Beneke:2001ev}

 \begin{equation}\label{H1}
\begin{split}
 H_i{(M_1M_2)}&=\frac{f_Bf_{M_1}}{2m_V\epsilon_{V}^*\cdot p_BF_0^{B\rightarrow M_1}(0)}\int_{0}^{1}\frac{d\xi}{\xi}\Phi_{B}{(\xi)}\int_{0}^{1}dx\int_{0}^{1}dy{\bigg[\frac{\Phi_{M_2}{(x)}\Phi_{M_1}{(y)}}{\bar{x}\bar{y}}+{r_\chi^{M_1}}\frac{\Phi_{M_2}{(x)}\Phi_{m_1}{(y)}}{x\bar{y}}\bigg]},
\end{split}
\end{equation}
for $i=1-4,9,10$,
\begin{equation}\label{H2}
\begin{split}
 H_i{(M_1M_2)}&=-\frac{f_Bf_{M_1}}{2m_V\epsilon_{V}^*\cdot p_BF_0^{B\rightarrow M_1}(0)}\int_{0}^{1}\frac{d\xi}{\xi}\Phi_{B}{(\xi)}\int_{0}^{1}dx\int_{0}^{1}dy{\bigg[\frac{\Phi_{M_2}{(x)}\Phi_{M_1}{(y)}}{{x}\bar{y}}+{r_\chi^{M_1}}\frac{\Phi_{M_2}{(x)}\Phi_{m_1}{(y)}}{\bar{x}\bar{y}}\bigg]},
\end{split}
\end{equation}
for $i=5,7$ and $H_i(M_1M_2)=0$ for $i=6,8$.

When $M_1M_2=SP,PS$ \cite{Beneke:2003zv,Cheng:2005nb},

\begin{equation}\label{H3}
\begin{split}
 H_i{(M_1M_2)}&=\frac{f_Bf_{M_1}}{f_{M_2}F_0^{B\rightarrow M_1}m_B^2}\int_{0}^{1}\frac{d\xi}{\xi}\Phi_{B}{(\xi)}\int_{0}^{1}dx\int_{0}^{1}dy{\bigg[\frac{\Phi_{M_2}{(x)}\Phi_{M_1}{(y)}}{\bar{x}\bar{y}}+{r_\chi^{M_1}}\frac{\Phi_{M_2}{(x)}\Phi_{m_1}{(y)}}{x\bar{y}}\bigg]},
\end{split}
\end{equation}
for $i=1-4,9,10$,
\begin{equation}\label{H4}
\begin{split}
 H_i{(M_1M_2)}&=-\frac{f_Bf_{M_1}}{f_{M_2}F_0^{B\rightarrow M_1}m_B^2}\int_{0}^{1}\frac{d\xi}{\xi}\Phi_{B}{(\xi)}\int_{0}^{1}dx\int_{0}^{1}dy{\bigg[\frac{\Phi_{M_2}{(x)}\Phi_{M_1}{(y)}}{{x}\bar{y}}+{r_\chi^{M_1}}\frac{\Phi_{M_2}{(x)}\Phi_{m_1}{(y)}}{\bar{x}\bar{y}}\bigg]},
\end{split}
\end{equation}
for $i=5,7$ and $H_i(M_1M_2)=0$ for $i=6,8$.
The coefficients of flavor operators $\alpha_i^p$ can be expressed in terms of the effective coefficients $a_i^p$ as
 \begin{equation}\label{alphai}
 \begin{split}
 \alpha_1{(M_1M_2)}&=a_1{(M_1M_2)},\\
\alpha_2{(M_1M_2)}&=a_2{(M_1M_2)},\\
\alpha_3^p{(M_1M_2)}&=\begin{cases}
a_3^p{(M_1M_2)}-a_5^p{(M_1M_2)}, \quad \text{if $M_1M_2=VP,SP$},\\
a_3^p{(M_1M_2)}+a_5^p{(M_1M_2)}, \quad \text{if $M_1M_2=PV,PS$},\\
\end{cases}\\
\alpha_4^p{(M_1M_2)}&=\begin{cases}
a_4^p{(M_1M_2)}+r_\chi^{M_2}a_6^p{(M_1M_2)}, \quad \text{if $M_1M_2=PV$},\\
a_4^p{(M_1M_2)}-r_\chi^{M_2}a_6^p{(M_1M_2)},\quad \text{if $M_1M_2=VP,PS,SP$},\\
\end{cases}\\
\alpha_{3,EW}^p{(M_1M_2)}&=\begin{cases}
a_9^p{(M_1M_2)}-a_7^p{(M_1M_2)},\quad \text{if $M_1M_2=VP,SP$},\\
a_9^p{(M_1M_2)}+a_7^p{(M_1M_2)},\quad \text{if $M_1M_2=PV,PS$},\\
\end{cases}\\
\alpha_{4,EW}^p{(M_1M_2)}&=\begin{cases}
 a_{10}^p{(M_1M_2)}+r_\chi^{M_2}a_8^p{(M_1M_2)},\quad \text{if $M_1M_2=PV$},\\
 a_{10}^p{(M_1M_2)}-r_\chi^{M_2}a_8^p{(M_1M_2)},\quad \text{if $M_1M_2=VP,PS,SP$},\\
 \end{cases}\\
  \end{split}
  \end{equation}
where the chiral factors $r_\chi$ are given by
\begin{equation}\label{rp}
\begin{split}
r_\chi^P{(\mu)}&=\frac{2m_P^2}{m_b(\mu)(m_1(\mu)+m_2(\mu))},\quad r_\chi^V{(\mu)}=\frac{2m_V^2}{m_b(\mu)}\frac{f_V^\perp(\mu)}{f_V},\\
r_\chi^S{(\mu)}&=\frac{2m_S^2}{m_b(\mu)}\frac{\bar{f}_S(\mu)}{f_S}=\frac{2m_S^2}{m_b(\mu)(m_1(\mu)+m_2(\mu))}.
\end{split}
\end{equation}

Weak annihilation contributions are described by the terms $b_i$ and $b_{i,EW}$, which have the following expressions:
\begin{equation}\label{b}
\begin{split}
&b_1=\frac{C_F}{N_c^2}C_1A_1^i, \quad b_2=\frac{C_F}{N_c^2}C_2A_1^i, \\
&b_3^p=\frac{C_F}{N_c^2}\bigg[C_3A_1^i+C_5(A_3^i+A_3^f)+N_cC_6A_3^f \bigg],\quad b_4^p=\frac{C_F}{N_c^2}\bigg[C_4A_1^i+C_6A_2^i \bigg], \\
&b_{3,EW}^p=\frac{C_F}{N_c^2}\bigg[C_9A_1^i+C_7(A_3^i+A_3^f)+N_cC_8A_3^f \bigg],\\
&b_{4,EW}^p=\frac{C_F}{N_c^2}\bigg[C_{10}A_1^i+C_8A_2^i \bigg],
\end{split}
 \end{equation}
where the subscripts 1, 2, 3 of $A_n^{i,f}(n=1,2,3)$ correspond to the possible Dirac structures $(V-A)(V-A)$, $(V-A)(V+A)$, and $(S-P)(S+P)$, respectively, and the superscripts $i$ and $f$ refer to gluon emission from the initial- and final-state quarks, respectively. Their explicit expressions are given by \cite{Beneke:2003zv,Cheng:2005nb}
\begin{equation}\label{Ai}
\begin{split}
 A_1^i&=\pi\alpha_s\int_0^1 dx dy\begin{cases}
 \bigg(\Phi_{M_2}(x)\Phi_{M_1}(y)\bigg[\frac{1}{y(1-x\bar{y})}+\frac{1}{\bar{x}^2y}\bigg]-r_\chi^{M_1}r_\chi^{M_2} \Phi_{m_2}(x)\Phi_{m_1}(y)\frac{2}{\bar{x}y}\bigg),\quad \text{for $M_1M_2=VP,PS,$}\\
 \bigg(\Phi_{M_2}(x)\Phi_{M_1}(y)\bigg[\frac{1}{y(1-x\bar{y})}+\frac{1}{\bar{x}^2y}\bigg]+r_\chi^{M_1}r_\chi^{M_2} \Phi_{m_2}(x)\Phi_{m_1}(y)\frac{2}{\bar{x}y}\bigg),\quad \text{for $M_1M_2=PV,SP,$}\\
 \end{cases}\\
 A_2^i&=\pi\alpha_s\int_0^1 dx dy\begin{cases}
 \bigg(-\Phi_{M_2}(x)\Phi_{M_1}(y)\bigg[\frac{1}{\bar{x}(1-x\bar{y})}+\frac{1}{\bar{x}y^2}\bigg]+r_\chi^{M_1}r_\chi^{M_2} \Phi_{m_2}(x)\Phi_{m_1}(y)\frac{2}{\bar{x}y}\bigg),\quad \text{for $M_1M_2=VP,PS,$}\\
 \bigg(-\Phi_{M_2}(x)\Phi_{M_1}(y)\bigg[\frac{1}{\bar{x}(1-x\bar{y})}+\frac{1}{\bar{x}y^2}\bigg]-r_\chi^{M_1}r_\chi^{M_2} \Phi_{m_2}(x)\Phi_{m_1}(y)\frac{2}{\bar{x}y}\bigg),\quad \text{for $M_1M_2=PV,SP,$}\\
 \end{cases}\\
 A_3^i&=\pi\alpha_s\int_0^1 dx dy\begin{cases}
 \bigg(r_\chi^{M_1}\Phi_{M_2}(x)\Phi_{m_1}(y)\frac{2\overline{y}}{\overline{x}y(1-x\bar{y})}+r_\chi^{M_2} \Phi_{M_1}(y)\Phi_{m_2}(x)\frac{2x}{\overline{x}y(1-x\bar{y})}\bigg),\quad \text{for $M_1M_2=VP,PS,$}\\
 \bigg(-r_\chi^{M_1}\Phi_{M_2}(x)\Phi_{m_1}(y)\frac{2\overline{y}}{\overline{x}y(1-x\bar{y})}+r_\chi^{M_2} \Phi_{M_1}(y)\Phi_{m_2}(x)\frac{2x}{\overline{x}y(1-x\bar{y})}\bigg),\quad \text{for $M_1M_2=PV,SP,$}\\
 \end{cases}\\
 A_3^f&=\pi\alpha_s\int_0^1 dx dy\begin{cases}
 \bigg(r_\chi^{M_1}\Phi_{M_2}(x)\Phi_{m_1}(y)\frac{2(1+\overline{x})}{\overline{x}^2y}-r_\chi^{M_2} \Phi_{M_1}(y)\Phi_{m_2}(x)\frac{2(1+y)}{\overline{x}y^2}\bigg),\quad \text{for $M_1M_2=VP,PS,$}\\
 \bigg(-r_\chi^{M_1}\Phi_{M_2}(x)\Phi_{m_1}(y)\frac{2(1+\overline{x})}{\overline{x}^2y}-r_\chi^{M_2} \Phi_{M_1}(y)\Phi_{m_2}(x)\frac{2(1+y)}{\overline{x}y^2}\bigg),\quad \text{for $M_1M_2=PV,SP,$}\\
 \end{cases}\\
 A_1^f&=A_2^f=0.\\
 \end{split}
  \end{equation}

The calculations of the hard-spectator interaction corrections involve the twist-3 distribution amplitude, $\Phi_{m_i}(x)$, which is non-zero as $x\rightarrow1$ in convolutions. These contributions lead to end-point divergences due to the non-zero end-point behavior of $\Phi_{m_i}$. One can  extract this divergence by defining a parameter $X_H$ through \cite{Beneke:2003zv}

\begin{equation}\label{XH}
\begin{split}
\int_0^1\frac{dx}{\bar{x}}\Phi_{m_i}(x)&=\Phi_{m_i}(1)\int_0^1\frac{dx}{\bar{x}}+\int_0^1\frac{dx}{\bar{x}}[\Phi_{m_i}(x)-\Phi_{m_i}(1)],\\
&\equiv \Phi_{m_i}(1)X_H^{M_iM_j}+\int_0^1\frac{dx}{[\bar{x}]_+}\Phi_{m_i}(x),\\
\end{split}
\end{equation}
with $\bar{x}=1-x$. Although the remaining integral is finite, the parameter $X_H^{M_i}$ representing the soft-gluon interaction with the spectator quark is unknown. It is expected that $X_H^{M_iM_j}\sim \mathrm{ln}(m_b/\mathrm{\Lambda_{QCD}})$ because it arises in a perturbative calculation of these soft interactions that are regulated in principle by a physical scale of order $\Lambda_{\mathrm{QCD}}$, however with a potentially complex coefficient, since multiple soft scatterings can introduce a strong-interaction phase. We encounter a similar end-point divergence issue in weak-annihilation processes, and can address it using the same approach as the power corrections for hard-spectator scatterings. The divergent subtractions are interpreted as

\begin{equation}\label{XA}
\int_0^1\frac{dx}{x}\rightarrow X_A^{M_iM_j},\quad \int_0^1dx\frac{\mathrm{ln}x}{x}\rightarrow-\frac{1}{2}(X_A^{M_iM_j})^2.\\
\end{equation}

The contributions from hard-scattering and weak-annihilation topologies, which are subleading in $1/m_b$, have yet to be systematically addressed in QCDF. However, they can be chirally enhanced and contribute sizable corrections to leading-order results. To parameterize QCD effects in $X_{H}^{M_iM_j}$ and $X_{A}^{M_iM_j}$, additional phenomenological parameters, $|\rho_k|$ and $\phi_k$, were introduced as in the following ($X_k$ could be $X_{H}^{M_iM_j}$ and $X_{A}^{M_iM_j}$):
\begin{equation}\label{XHA}
\begin{split}
\begin{cases}
& X_k=(1+\rho_k)\ln\frac{m_B}{\Lambda_h},  \\
& \rho_k=|\rho_k| e^{i\phi_k}, \\
\end{cases}\quad \text{with}\quad \phi_k\in[0,2\pi],\quad\Lambda_h= 0.5\mathrm{GeV}.
\end{split}
 \end{equation}

In general, the total amplitude of the $B$ meson three-body decay $B\rightarrow M_1M_2M_3$ ($m_1$, $m_2$, $m_3$ being the masses of $M_1$, $M_2$, and $M_3$, respectively) can be defined as a coherent sum of $N$ different components, each of which can be described by a parameterized function $F_j$ ($j=1,\cdot\cdot\cdot,N$) representing the contribution of the $j$-th intermediate resonance state,
\begin{equation} \label{MT}
\mathcal{M}(m_{12}^2,m_{23}^2)=\sum_j^N e^{i\delta_j}F_j(m_{12}^2,m_{23}^2),
\end{equation}
where $\delta_j$ is a strong phase, $m_{ik}$ is the invariant mass of  the system of $M_i$ and $M_k$ ($i, k=1,2,3$), and the function $F_j$ contains not only weak interaction but also strong interaction contributions, which, for a resonance decaying into $M_1$ and $M_2$, is parametrized as
\begin{equation} \label{Fj}
F_j(m_{12}^2,m_{23}^2)=M_{wj}(m_{12}^2,m_{23}^2)\cdot R(m_{12})\cdot M_{sj}(\vec{p},\vec{q}),\\
\end{equation}
where $M_{wj}(m_{12}^2,m_{23}^2)$ and $M_{sj}(\vec{p},\vec{q})$ are the decay amplitudes for the weak decay of the $B$ meson and the strong decay of the intermediate resonance,
respectively, $R(m_{12})$  describes the propagator of this intermediate resonance, $\vec{p}$ represents the momentum of $M_3$, $\vec{q}$ denotes the momentum of $M_1$, where both $\vec{p}$ and $\vec{q}$ are evaluated in the rest frame of the $M_1M_2$ system.

\section{THE DEFINITIONS OF THE FBA AND THE FB-CPA}
\begin{figure}[ht]
\centerline{\includegraphics[width=0.65\textwidth]{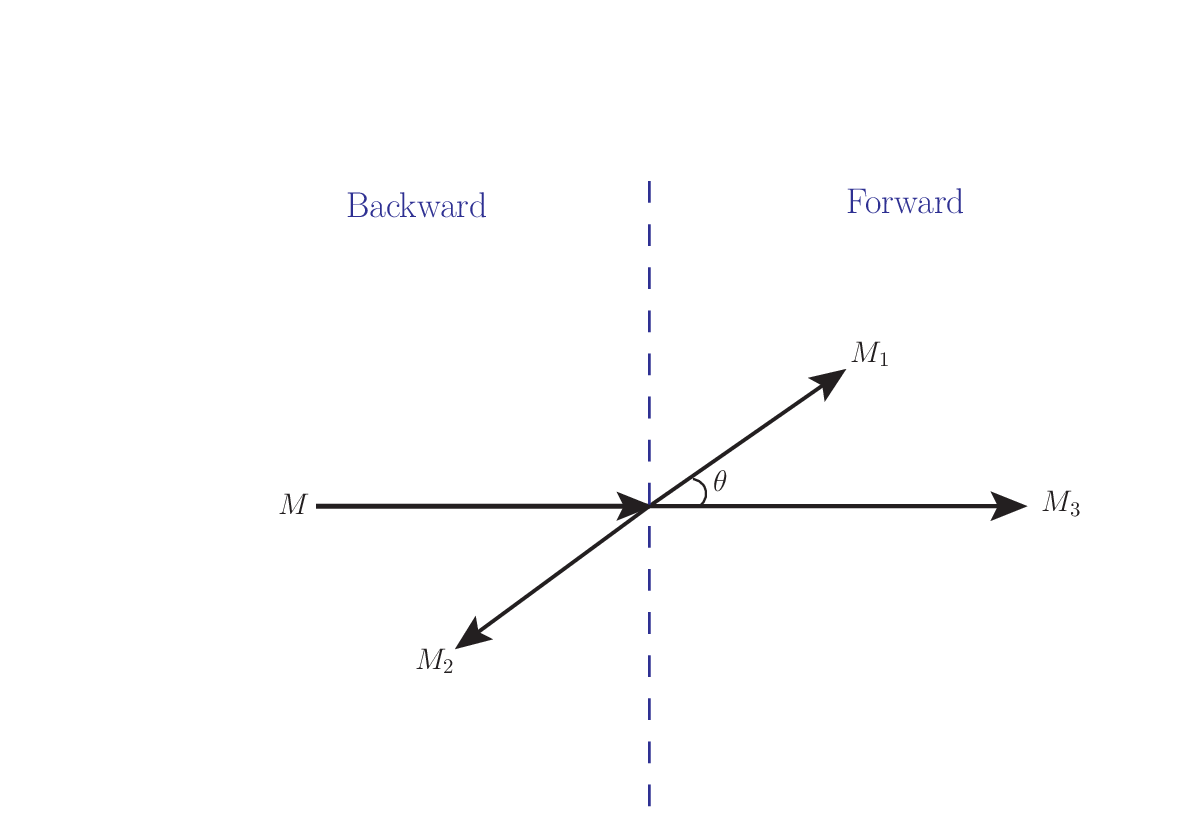}}
\caption{The definition of $\theta$ in the c.m. frame of the $M_1M_2$ system ($M$ is chosen to be the $B$ meson).}
\label{M1M2frame}
\end{figure}

The localized $CP$ asymmetry, $\mathcal{A}^L_{CP}$, has received extensive attention from both experimental and theoretical perspectives for a three-body decay $B\rightarrow M_1M_2M_3$ as shown in Fig. {\ref{M1M2frame}} (the conjugate process is  $\bar{B}\rightarrow \bar{M}_1\bar{M}_2\bar{M}_3$ ), with $M_i$ $(i=1,2,3)$ being light mesons.
Take $B^-$ as an example, $\mathcal{A}^L_{CP}$ is a function of the $B^-$ and $B^+$ decay event yields ($N_{B^-}$ and $N_{B^+}$) as,
\begin{equation} \label{AFBCP}
\mathcal{A}^L_{CP}=\frac{N_{B^-}(\mathrm{Forward}+\mathrm{Backward})-N_{B^+}(\mathrm{Forward}+\mathrm{Backward})}{N_{B^-}(\mathrm{Forward}
+\mathrm{Backward})+N_{B^+}(\mathrm{Forward}+\mathrm{Backward})}\\,
\end{equation}
where the ``Forward" describes the events in which $M_1$ moves forward with respect to the initial state $B$ in the c.m. frame of the $M_1M_2$ system, marked by $\theta<\frac{\pi}{2}$, and conversely, the ``Backward" describes events corresponding to the backward motion of $M_1$ with $\theta>\frac{\pi}{2}$. The event yield of each bin $i$, $N_{B^\pm,i}$ has the following form:

\begin{equation}\label{NBi}
\begin{split}
N_{B^\pm,i}=\eta\int_{\mathrm{\mathrm{cos\theta\gtrless0}}}R|\mathcal{M}_{B^\pm\rightarrow M_1^\pm M_2^+M_3^-}|^2_{m_{12}=m_{12,i}}d\mathrm{cos\theta},\\
\end{split}
\end{equation}
where $\eta$ is the ratio of the total number of events to the total decay width, and here we treat it as an unknown parameter, $R$ is the phase-space factor with the form $R=\sqrt{(m_{12}^2-4m_{M_2}^2)[m_B^2-(m_{12}-m_{M_3})^2][m_B^2-(m_{12}+m_{M_3})^2]}$ \footnote{Since we integrate over the phase space angle $\theta$ instead of the invariant mass square $m_{23}^2$, a factor $DF$, $(m^2_{23,\mathrm{max}}-m^2_{23,\mathrm{min}})/2$ appears due to the relation $\cos\theta=\frac{m^2_{23}-(m^2_{23,\mathrm{max}}+m^2_{23,\mathrm{min}})/2}{(m^2_{23,\mathrm{max}}-m^2_{23,\mathrm{min}})/2}$. By applying the boundary conditions of the Dalitz plot and absorbing the denominator of $DF$ into the amplitude, $DF$ becomes $R$, where $R=\sqrt{(m_{12}^2-4m_{M_2}^2)[m_B^2-(m_{12}-m_{M_3})^2][m_B^2-(m_{12}+m_{M_3})^2]}$.}, and $m_{12,i}$ is the average value of $m_{12}$ in each bin $i$.

One can introduce a new physical quantity FB-CPA to study the $CP$ asymmetry in the multi-body decay processes of $B$ mesons. It can provide a good approach for isolating the interference effects between nearby intermediate resonances. The specific form of FB-CPA is as follows: \cite{Wei:2022zuf,Hu:2022eql,Zhang:2021zhr}
\begin{equation} \label{AFBCP}
\mathcal{A}_{FB}^{CP}=\frac{1}{2}(\mathcal{A}_{FB}-\bar{\mathcal{A}}_{FB}),
\end{equation}
where $\mathcal{A}_{FB}$ and $\bar{\mathcal{A}}_{FB}$ are the forward-backward asymmetries for $B\rightarrow M_1M_2M_3$ and it's conjugate process, respectively. Generating a significant forward-backward asymmetry is challenging when only even or odd spin resonanaces are present. However, it becomes relatively easy to achieve a substantial FBA when considering the interference contributions from nearby spin-odd (and even) resonances. In the forward and backward hemispheres, the  specific definition of FBA in Eq. (\ref{AFBCP}) is \cite{Gutsche:2013pp}

\begin{equation} \label{AFB1}
\begin{split}
\mathcal{A}_{FB}&=\frac{N_{B^\mp}(\mathrm{Forward})-N_{B^\mp}(\mathrm{Backward})}{N_{B^\mp}(\mathrm{Forward})+N_{B^\mp}(\mathrm{Backward})},\\
&=\frac{N_{B^\mp}(\mathrm{\cos\theta>0})-N_{B^\mp}(\mathrm{\cos\theta<0})}{N_{B^\mp}(\mathrm{\cos\theta>0})+N_{B^\mp}(\mathrm{\cos\theta<0})}.\\
\end{split}
\end{equation}

Theoretically, when we focus on the phase space near the intermediate resonant state $X$,  the corresponding decay will be dominated by the cascade decay $B\rightarrow XM_3\rightarrow M_1M_2M_3$. Then, this part of phase space can be further divided into two parts according to whether $\theta$ is larger or smaller than $\pi/2$ (that is, $\cos\theta<0$ or $\cos\theta>0$ ). Thus, the forward-backward asymmetry, $\mathcal{A}_{FB}$, can be described in these two different phase spaces as \cite{Zhang:2021zhr}
\begin{equation} \label{AFB2}
\mathcal{A}_{FB}=\frac{\int_0^1\langle|\mathcal{M}|^2\rangle d\cos\theta-\int_{-1}^0\langle|\mathcal{M}|^2\rangle d\cos\theta}{\int_{-1}^1\langle|\mathcal{M}|^2\rangle d\cos\theta},\\
\end{equation}
where $\langle|\mathcal{M}|^2\rangle$ takes the specific form $\frac{1}{2}\int_g^h (m^2_{23,\mathrm{max}}-m^2_{23,\mathrm{min}})|\mathcal{M}|^2dm^2_{12}$, with $[g,h]$ being the integration interval of $m_{12}$ on which we focus, i.e., the region $(m_X-\Gamma_X)<m_{12}<(m_X+\Gamma_X)$, where $m_X$ and $\Gamma_X$ are the mass and the decay width of $X$, respectively, $m^2_{23,\mathrm{min}(\mathrm{max})}$ is the minimum (maximum) value of $m_{23}$ for fixed $m_{12}$, $\cos\theta=\frac{\vec{p}_1\cdot\vec{p}_3}{|\vec{p}_1||\vec{p}_3|}$ in the c.m. frame of the $M_1M_2$ system, where $\vec{p}_i$ is the momentum of $M_j$. After a derivation, one can get the relationship between $\cos\theta$ and $m_{23}$ as $\cos\theta=\frac{m^2_{23}-(m^2_{23,\mathrm{max}}+m^2_{23,\mathrm{min}})/2}{(m^2_{23,\mathrm{max}}-m^2_{23,\mathrm{min}})/2}$. In terms of partial waves, the decay amplitude $\mathcal{M}$ can be expressed as $\mathcal{M}=\sum\limits_la_lP_l(\cos\theta)$, then the FBA becomes $\mathcal{A}_{FB}=\frac{2}{\Sigma_j[\langle|a_j|^2\rangle/(2j+1)]}\sum\limits_{\substack{\mathrm{even}\,l\\ \mathrm{odd}\,k}}f_{lk}\mathfrak{R}(\langle a_la_k^*\rangle)$ with $f_{lk}=\int_0^1P_lP_kd\cos\theta=\frac{(-1)^{(l+k+1)/2}l!k!}{2^{l+k-l}(l-k)(l+k+1)[(l/2)!]^2{[(k-1)/2]!}^2}$, where the numerator only includes the interference term between even- and odd- waves, implying that in order to obtain relatively large FBA, the interference between even- and odd- waves must be considered \cite{Wei:2022zuf,Hu:2022eql}.

\section{Numerical results}
Let us clarify the input parameters before presenting the numerical results. The values of the Wolfenstein parameters are given as \cite{Workman:2022ynf}
\begin{equation} \label{CKMm}
A=0.790^{+0.017}_{-0.012},\quad \lambda=0.22650^{+0.00048}_{-0.00048},\quad\bar{\rho}=0.141^{+0.016}_{-0.017},\quad \bar{\eta}=0.357^{+0.011}_{-0.011}.\\
\end{equation}

For the quark masses, we use \cite{Workman:2022ynf}
\begin{equation} \label{mass}
\begin{split}
 &m_s(\mu)/m_q(\mu)=(0.0273^{+0.0007}_{-0.0013})\times10^3,\quad m_s(2\mathrm{GeV})=0.093^{+0.011}_{-0.005} \mathrm{GeV},\quad m_b(m_b)=4.18^{+0.03}_{-0.02}\mathrm{GeV},  \\
 &m_c=1.67\pm0.07\mathrm{GeV},\quad m_b=4.78\pm0.06\mathrm{GeV},\quad m_t=172.76\pm0.30\mathrm{GeV},\quad m_q=(m_u+m_d)/2.\\
\end{split}
\end{equation}

The nonperturbative inputs used in this work include decay constants (in units of GeV), Gegenbauer moments and form factors for the pseudoscalar and vector mesons (for the scalars see Table \ref{dcmo}), we take \cite{ParticleDataGroup:2024cfk,Ball:2004rg,Al-Haydari:2009kal,Qi:2018syl}
\begin{equation} \label{decay constants form factors}
\begin{split}
&f_{\pi}=0.130,\quad f_{K}=0.156, \quad f_{B}= 0.190,\quad f_{\rho(770)^0}=0.216\pm0.003, \quad f_{\rho(770)^0}^\perp=0.165\pm0.009,\\
&\alpha_1^{\rho(770)^0}=0,\quad \alpha_2^{\rho(770)^0}=0.15\pm0.07,\quad \alpha_{1,\perp}^{\rho(770)^0}=0, \quad \alpha_{2,\perp}^{\rho(770)^0}=0.14\pm0.06,\\
&F_0^{B\rightarrow \pi}(0)=0.25\pm0.03,\quad F_0^{B\rightarrow K}(0)=0.35\pm0.04, \quad A_0^{B\rightarrow \rho}(0)= 0.303\pm0.029. \\
\end{split}
\end{equation}

\begin{figure}[htbp]
\centering
\subfigure[]{
\includegraphics[width=3.3in]{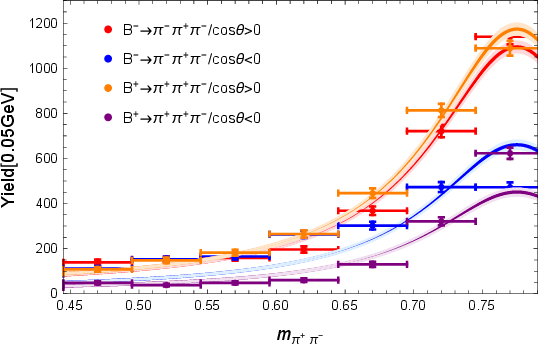}}
\subfigure[]{
\includegraphics[width=3.3in]{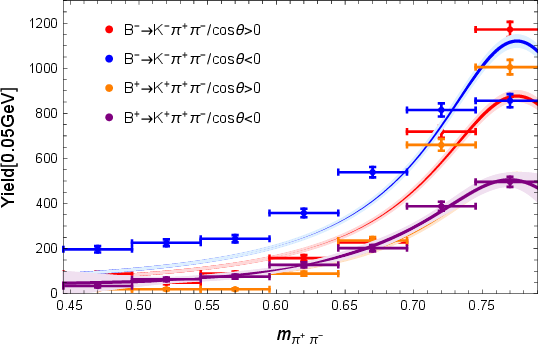}}
\caption{The fitted curves for the (a) $B^\pm\rightarrow\pi^\pm\pi^+\pi^-$  and (b) $B^\pm\rightarrow K^\pm\pi^+\pi^-$ decays, the event yields are from the data of LHCb \cite{LHCb:2014mir}. The contribution from the $\rho(770)^0$ are only considered in our theoretical calculations.}
\label{Theoevent}
\end{figure}

In this work, we focus on the $B^\pm\rightarrow \pi^\pm\pi^+\pi^-$ and $B^\pm\rightarrow K^\pm\pi^+\pi^-$ decays. Following the general framework in Sect. ${\mathrm{\uppercase\expandafter{\romannumeral3}}}$, before exploring $CP$ asymmetry observables caused by the interference between $\rho(770)^0$ and $f_0(500)$ intermediate resonances, such as FBA-CPAs and LCPAs, one should pay attention to the event yield distributions, as they are served as crucial links between theory and experiment studies. Prior to considering the contributions from the $\rho(770)^0$ and $f_0(500)$, we first investigate the event yield distributions with only the presence of the $\rho(770)^0$ using Eq. (\ref{NBi}) theoretically.  For the end-point divergence parameters of $B\rightarrow VP$  and $B\rightarrow PV$ decays, we adopt the $|\rho^{PV}_k|\approx0.87$, $|\rho^{VP}_k|\approx1.07$, $\phi^{PV}_k\approx-30^\circ$ and  $\phi^{VP}_k\approx-70^\circ$ as in Ref. \cite{Cheng:2009cn}, and assign an error of $\pm0.1$ to $\rho ^{VP/PV}_k$ and $\pm20^\circ$ to $\phi^{VP/PV}_k$, thus the unknown parameters $\eta$ is left. The experimental event yields in each bin $i$ are listed in Table. \ref{tabN}, which are obtained from the data of LHCb \cite{LHCb:2014mir}. With respect to the mass of $\pi\pi$ pair, the interval $0.445\mathrm{GeV}<m_{\pi\pi}<0.795\mathrm{GeV}$ is selected. Although both the $K\pi$ and $\pi\pi$ systems have impacts on the decay rates within this small in the Dalitz plots, the influence of the  $K\pi$ resonance is significantly weakened when studying the event yields. Therefore, we can safely consider the effects dominated by the $\rho(770)^0$ and $f_0(500)$ resonances \footnote{Let's take the $K_0^*(1430)$ resonance as an example, the ratio of the event numbers generated by the $K_0^*(1430)$  can be expressed as $\varpi=\frac{\Gamma_{\rho(770)^0}^2}{\Delta_{s_{\pi\pi}}}\frac{Br(B\rightarrow K_0^*(1430)\pi )}{Br(B\rightarrow \rho(770)^0K )}$ in the overlap of the $\rho(770)^0$ band and the $K_0^*(1430)$ band, where $\Delta_{s_{\pi\pi}}$ is the difference between the maximum value $s_{\pi\pi,\mathrm{max}}$ and the minimum value $s_{\pi\pi,\mathrm{min}}$ of the invariant mass square of the $\pi\pi$ pair when $m_{K\pi}=m_{K_0^*(1430)}$ in the Dalitz plot, that is $\Delta_{s_{\pi\pi}}=s_{\pi\pi,\mathrm{max}}-s_{\pi\pi,\mathrm{min}}$. One can get $\frac{\Gamma_{\rho(770)^0}^2}{\Delta_{s_{\pi\pi}}}\thickapprox 10^{-3}$, $\frac{Br(B\rightarrow K_0^*(1430)\pi )}{Br(B\rightarrow \rho(770)^0K )}=10$  and then $\varpi=10^{-2}$ (with $m_\pi=0.140$ GeV, $m_K=0.494$ GeV, $m_{\rho(770)^0}=0.775$ GeV, $\Gamma_{\rho(770)^0}=0.150$ GeV, $m_{K_0^*(1430)}=1.425$ GeV  and $\Gamma_{K_0^*(1430)}=0.270 $ GeV). Therefore, the contribution from the $K_0^*(1430)$ resonance is negligible.}. By general fitting with $\chi^2/\mathrm{d.o.f.}=37.30$ and $\eta=7.67\mathrm{GeV}^{-1}$, one get the related theoretical results which are displayed in Fig. \ref{Theoevent} and comparing with the experimental data. It is noted that there is a significant discrepancy between the theoretical and experimental results from Fig. \ref{Theoevent}, thus we discard this scenario and take into account the aforementioned interference between $\rho(770)^0$ and $f_0(500)$ .

\begin{table}[tb]
\renewcommand{\arraystretch}{1.2}
\centering
\caption{Form factors of $B\rightarrow S$ transitions at $q^2=0$ \cite{Cheng:2005nb}, the scalar decay constants $\bar{f}_s$ (in units of MeV) and the Gegenbauer moments at the scale $\mu=1\mathrm{GeV}$ \cite{Cheng:2013fba,Cheng:2020hyj,Cheng:2007st,Cheng:2022ysn}.}
\begin{tabular*}{\textwidth}{@{\extracolsep{\fill}}cc|ccccc}
\hline
\hline
 F &F(0) &Scalar & $\bar{f}_s$  & $B_1$       & $B_3$     \\
\hline
 $F_0^{Bf_0(500)}$&0.25 &$f_0(500)$  & $350\pm20$($\bar{f}^n_{f_0}$)\cite{Cheng:2020hyj}     &$-0.42\pm0.07$   &$0.01\pm0.04$\\
& &  & $-107\pm6$($\bar{f}^s_{f_0}$)     &$-0.42\pm0.07$   &$0.01\pm0.04$\\
 $F_0^{Bf_0(980)}$&0.25 &$f_0(980)$  & $113\pm6$($\bar{f}^n_{f_0}$)     & $-0.78\pm0.08$        & $0.02\pm0.07$           \\
 & &  & $370\pm20$ ($\bar{f}^s_{f_0}$)\cite{Cheng:2020hyj}    & $-0.78\pm0.08$        & $0.02\pm0.07$           \\
$F_0^{Ba_0(980)}$&0.25 &$a_0(980)$  & $365\pm20$    &$-0.93\pm0.10$      &$0.14\pm0.08$                 \\
$F_0^{Ba_0(1450)}$& 0.26&$a_0(1450)$  &  $460\pm50$   &$0.89\pm0.20$                       &$-1.38\pm0.18$\\
$F_0^{BK^{*}_0(1430)}$&0.26 &$K^{*}_0(1430)$ & $-300\pm30$    &$0.58\pm0.07$                       &$-1.20\pm0.08$\\
\hline
\hline
\end{tabular*}\label{dcmo}
\end{table}

\begin{table}[h!]
  \begin{center}
        \caption{The event yields for $B^\pm\rightarrow\pi^\pm\pi^+\pi^-$ and $B^\pm\rightarrow K^\pm\pi^+\pi^-$ decays with $m_{\pi\pi}$ ranging from 0.445 GeV to 0.795 GeV based on Figs. 4 and 5 in Ref. \cite{LHCb:2014mir}, where $N_i(\cos\theta\gtrless0)$ and $\bar{N}_i(\cos\theta\gtrless0)$ for $B^-\rightarrow\pi^- (K^-)\pi^+\pi^-$ and $B^+\rightarrow\pi^+ (K^+)\pi^+\pi^-$ decays, respectively, and the uncertainties are statistical only.
      }
  \begin{tabular}{c|cccc|cccc}
    \hline\hline
    \multirow{2}{*}{Bin (GeV)} &\multicolumn{4}{c|}{$B^\pm\rightarrow\pi^\pm\pi^+\pi^-$} & \multicolumn{4}{c}{$B^\pm\rightarrow K^\pm\pi^+\pi^-$}  \\
    \cline{2-9}
     & $N_i(\cos\theta\!\!>\!\!0)$ & $N_i(\cos\theta\!\!<\!\!0)$  & $\overline{N}_i(\cos\theta\!\!>\!\!0)$ & $\overline{N}_i(\cos\theta\!\!<\!\!0)$ & $N_i(\cos\theta\!\!>\!\!0)$ & $N_i(\cos\theta\!\!<\!\!0)$  & $\overline{N}_i(\cos\theta\!\!>\!\!0)$ & $\overline{N}_i(\cos\theta\!\!<\!\!0)$ \\
    \hline
    {0.445-0.495} &  $139\pm12$   & $112\pm11$    &  $108\pm10$   &  $48\pm7$  &  $89\pm9$   & $197\pm14$   &  $20\pm4$  &$34\pm6$\\
        \hline
    {0.495-0.545} &  $153\pm12$   & $152\pm12$    &  $148\pm12$   & $38\pm6$   & $49\pm7$    & $226\pm15$   &  $20\pm4$  &$64\pm8$\\
        \hline
    {0.545-0.595} &  $159\pm13$  & $165\pm13$    &  $182\pm13$   &  $48\pm7$  & $89\pm9$    & $244\pm16$   &  $20\pm4$ &$76\pm9$ \\
        \hline
    {0.595-0.645} &  $196\pm14$   & $263\pm16$    &  $265\pm16$   & $60\pm8$   & $158\pm13$    & $358\pm19$   & $89\pm9$   &$128\pm11$\\
        \hline
    {0.645-0.695} &  $368\pm19$   & $302\pm17$    &  $446\pm21$   & $130\pm11$   & $227\pm15$    & $539\pm23$   & $236\pm15$  &$202\pm14$ \\
        \hline
    {0.695-0.745} &  $721\pm27$   & $473\pm22$    &  $813\pm29$   & $321\pm18$   & $719\pm27$    & $815\pm29$   &  $660\pm26$  &$388\pm20$ \\
    \hline
    {0.745-0.795} &  $1140\pm34$   & $472\pm22$    &  $1089\pm33$   & $623\pm25$   & $1172\pm34$    & $856\pm29$   &  $1005\pm32$  &$496\pm22$ \\
  \hline\hline
  \end{tabular}\label{tabN}
  \end{center}
\end{table}

\begin{figure}[htbp]
\centering
\subfigure[]{
\includegraphics[width=3.3in]{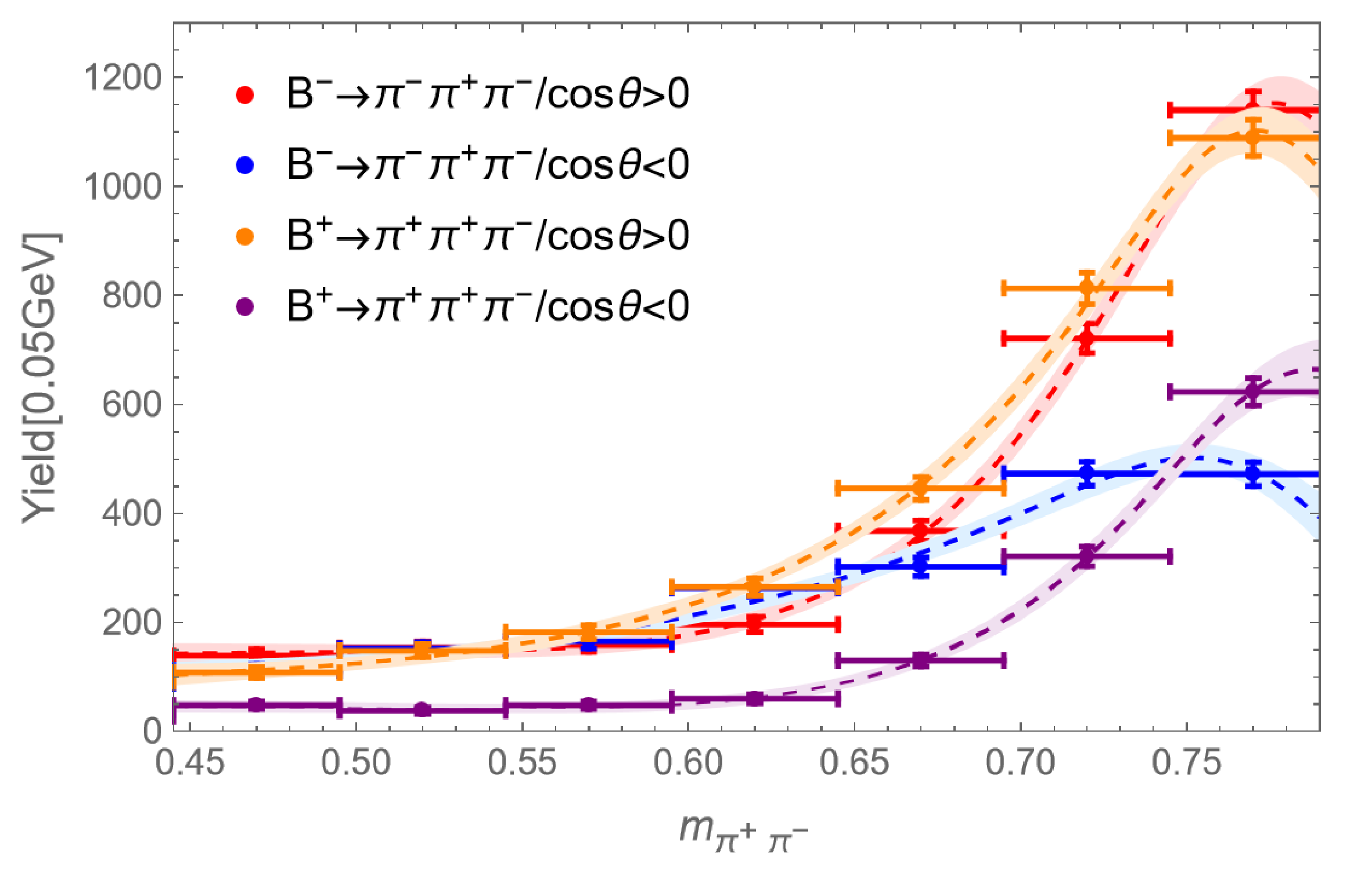}}
\subfigure[]{
\includegraphics[width=3.3in]{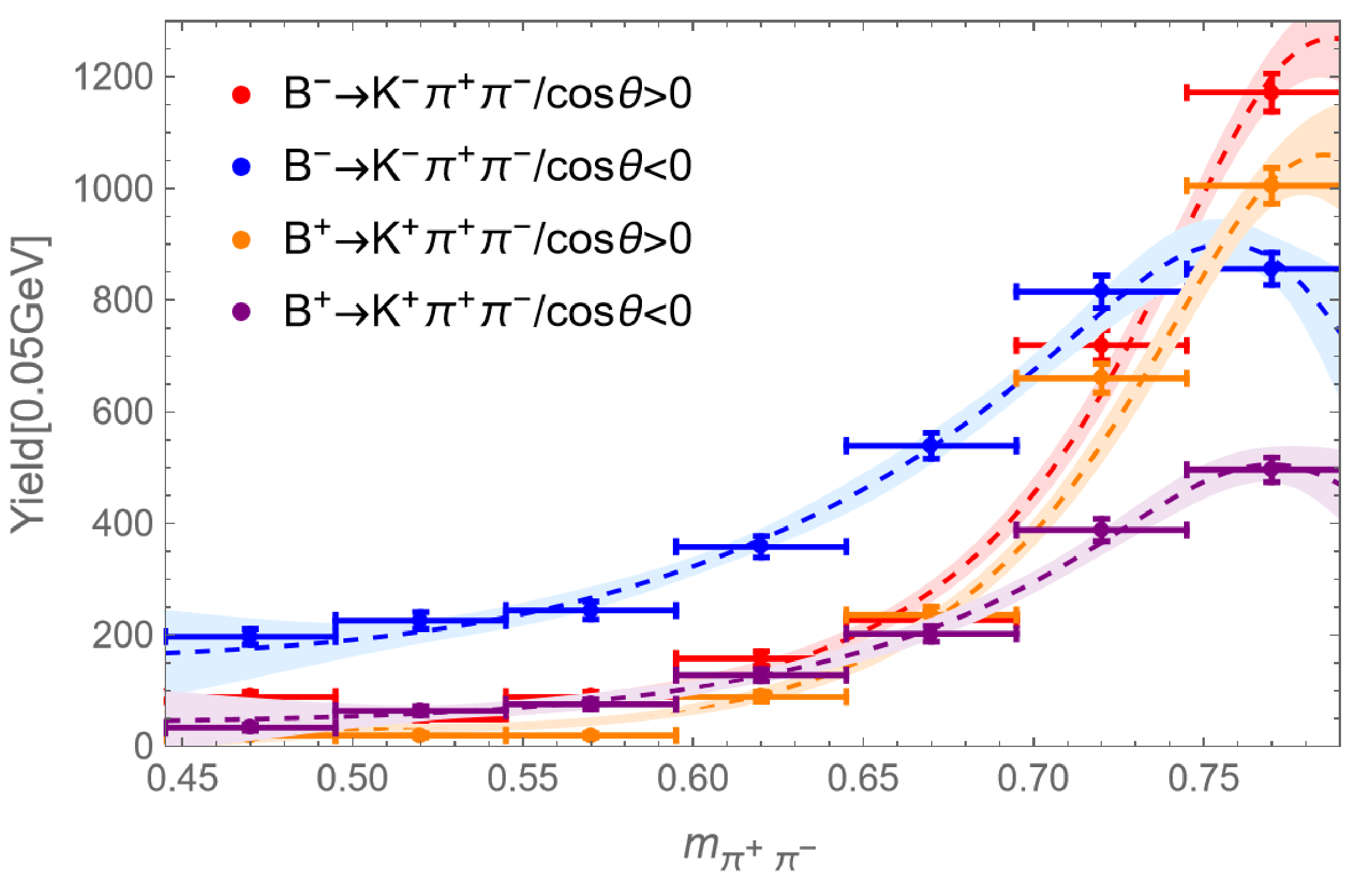}}
\caption{The fitted curves for the (a) $B^\pm\rightarrow\pi^\pm\pi^+\pi^-$  and (b) $B^\pm\rightarrow K^\pm\pi^+\pi^-$ decays, the event yields are from the data of LHCb \cite{LHCb:2014mir}. The contribution from the $\rho(770)^0$ and $f_0(500)$ are considered in our theoretical calculations.}
\label{NfitFIG}
\end{figure}

The quark structure of the light scalar mesons below or near 1 GeV has been quite controversial, though it is commonly believed that the $f_0(500)$, $f_0(980)$ and $a_0^0(980)$ are primarily four-quark or molecular states. In fact, it is difficult to provide quantitative predictions for $B\rightarrow SP/PS$ decays based on the four-quark model, as this not only involves unknown form factors and decay constants beyond the traditional quark model, but also includes difficult-to-estimate additional non-factorizable contributions. Therefore, in order to facilitate the application of QCDF, this work is based on the $q\bar{q}$  assignment for the scalar mesons for research. When consider the interference between $\rho(770)^0$ and $f_0(500)$, there exist $|\rho_k|$ and $\phi_k$ (especially $|\rho_k|$ ) in the expression of $X_k$ in Eq. (\ref{XHA}), which are utterly unknown from the first principles of QCD dynamics. We treat the parameters $|\rho_k|$ and $\phi_k$ for the $B\rightarrow SP/PS$ processes as distinct unknown parameters, $|\rho_k^{SP}|$, $|\rho_k^{PS}|$, $\phi_k^{SP}$ and $\phi_k^{PS}$, respectively \footnote{There are quite abundant experimental data and theoretical studies for the $B\rightarrow PP$ or $VP(PV)$ decays compared with the $B\rightarrow SP(PS)$ decays. Therefore,  we have adopted the end-point divergence parameters for the $B\rightarrow VP(PV)$ decays as in Ref. \cite{Cheng:2009cn}, while treating the end-point divergence parameters for the $B\rightarrow SP(PS)$ decay channel as unknowns.}.  Regarding the strong phases associated with $\rho(770)^0$ and $f_0(500)$ resonances in Eq. (\ref{MT}), we can make the following convention: taking $\delta_{\rho(770)^0}=0$, and $\delta_{f_0(500)}$ as an unknown quantity, i.e., the relative strong phase. Thus, when dealing with the amplitudes caused by $\rho(770)^0$, $f_0(500)$ and their interference, there six free parameters left, the relative strong phase $\delta_{f_0(500)}$, the four variables mentioned in the analysis of  the end-point divergences and the introduced parameter $\eta$ in Eq. (\ref{NBi}).  The event yields may be more suitable than other physical quantities to extract these six unknown parameters, because they are the most direct and raw observables presented by experiments.  The required decay amplitudes and propagators are listed in Appendices A and B, respectively. We have performed simultaneous fitting of the theoretical event yields in each bin $i$ with the six variables in Eq. (\ref{NBi}) to the experimental results of the $B^\pm\rightarrow \pi^\pm\pi^+\pi^-$ and $B^\pm\rightarrow K^\pm\pi^+\pi^-$ decay processes which are listed in Table \ref{tabN}, and the corresponding fitted curves are shown in Fig. \ref{NfitFIG}. The corresponding goodness-of-fit parameter is $\chi^2/\mathrm{d.o.f.}=3.79$, and the associated values of the fitting parameters are
\begin{equation}\label{rhophi}
\begin{split}
\begin{cases}
& |\rho_k^{SP}|=3.29\pm1.01 ,\quad |\phi_k^{SP}|=0.85\pm0.31;\\
& |\rho_k^{PS}|=2.33\pm0.73,\quad |\phi_k^{PS}|=0.15\pm0.08 ;\\
& \delta_{f_0(500)}=0.85\pm0.39,\quad \eta=(2.47\pm0.83)\mathrm{GeV}^{-1}.\\
\end{cases}
\end{split}
\end{equation}

By comparing Figs. 2 and 3,  we can see that the theoretical results can fit well with the experimental data after incorporating the contribution from the $f_0(500)$ resonance. However, the results of $|\rho_k^{SP}|$ and $|\rho_k^{PS}|$ in Eq. (\ref{rhophi}) are contradictory to the traditional choice of $\rho_k<1$. In general, $X_k$ in Eq. (\ref{XHA}) is expected to be around $\ln( m_b/\Lambda_{\mathrm{QCD}})$ which is regulated by a physical scale of order $\Lambda_{\mathrm{QCD}}$. However, due to the lackness of  a complete understanding of the degrees of freedom and their interactions below this scale, the numerical value of the complex parameter $X_k$ is unknown and model dependent, thus in QCDF a conservative convention $|\rho_k|<1$ is adopted and it is assumed that $x_k$ it is universal for all decay processes. The large value of $|\rho_k|$ implies that the contributions of hard-scattering and weak-annihilation which are formally $1/m_b$ power suppressed,  are not only nonnegligible but also significant, especially for the flavor-changing neutral-current dominated and pure annihilation decays. There are many studies suggesting the existence of unexpected large annihilation contributions by CDF, Belle and LHCb in $B_s\rightarrow \pi\pi$ and $B_d\rightarrow KK$ decays \cite{CDF:2011who,Belle:2012dmz,LHCb:2012ihl}, and the possible nonnegligible large weak-annihilation contributions were also noticed in the pQCD  approach \cite{Xiao:2011tx}.  In other words, if $|\rho_k|<1$ is confirmed, another potential explanation for our findings could be that the two-quark model interpretation of the scalar mesons may not be applicable. Of course, this  would also pose new challenges for QCDF, as the incorporation of scalar mesons in the four-quark model would introduce extra non-perturbative parameters, potentially making QCDF powerless in this situation.

It should be noted that we adopt the same set of endpoint divergence parameters for both light and heavy scalar mesons for an attempt. Using the fitted results of $|\rho_k^{SP}|$ and $|\rho_k^{PS}|$ in Eq. (\ref{rhophi}), we calculate the branching ratios and $CP$ violating asymmetries for the $B$ decays involving other scalar mesons, which are listed in Tables \ref{branching fraction} and \ref{Dcp}, respectively, and compared them with other theoretical and experimental results to investigate whether the large value of $|\rho_k^{SP}|$ and $|\rho_k^{PS}|$ is reasonable.   The nonperturbative inputs include decay constants, the Gegenbauer moments and form factors, which are summarized in Table. \ref{dcmo}, respectively.  It is essential to clarify that all the form factors are evaluated at $q^2=0$ due to the smallness of $m_\pi^2$ and $m_K^2$ compared with $m_B^2$ \cite{Beneke:2003zv}. We adopt the Flatt$\acute{\mathrm{e}}$ model and LASS model to deal with the propagators of the $f_0(980)$  and $K_0^*(1430)$, and for other scalar mesons we use the Breit-Wigner forms.

In the naive $q\bar{q}$ quark model,  $f_0(980)$ is purely an $s\bar{s}$ state, and $f_0(500)$ is a $n\bar{n}$ state with $n\bar{n}=(u\bar{u}+d\bar{d})/\sqrt{2}$. However, there
exist some experimental evidences clearly showing the existence between the strange and nonstrange quarks in $f_0(980)$. Therefore, isoscalars $f_0(980)$ and $f_0(500)$ should mixing with each other:
\begin{equation}\label{ffmixing}
\begin{split}
\left(
\begin{array}{cc}
|f_0(980)\rangle\\
 |f_0(500)\rangle\\
\end{array}
\right)=\left(
\begin{array}{cccc}
\cos\theta\quad \sin\theta\\
-\sin\theta\quad \cos\theta\\
\end{array}
\right)
\left(
\begin{array}{cc}
|s\bar{s}\rangle\\
|n\bar{n}\rangle\\
\end{array}
\right),
\end{split}
\end{equation}
where $\theta$ is the mixing angle between $f_0(980)$ and $f_0(500)$. LHCb found $\theta$ lies in the range of $0^\circ<\theta<30^\circ$ when measuring the upper limit on the branching ratio product $\mathcal{B}({\bar{B^0}\rightarrow J/\psi f_0(980)\rightarrow J/\psi \pi\pi})$ \cite{LHCb:2013dkk}. Besides, when explaining the large rates of $f_0(980)K$ and $f_0(980)K^*$ modes, it was found $\theta$ is around $20^\circ$ in Ref. \cite{Cheng:2013fba}, with the specific value being $\theta=17^\circ$.  In our calculations,  we also adopt this value for studying $B^-\rightarrow\pi^-(K^-)f_0(500)$ and $B^-\rightarrow\pi^-(K^-)f_0(980)$ decays. Using the QCD sum rule method  and $\theta=17^\circ$ \cite{Cheng:2005nb}, one can get the scalar decay constants with different quark components, which are listed in Table \ref{dcmo}. The superscript $n$ and $s$ of the decay constants  correspond to the nonstrange and strange quark contents of $f_0(500)$ and $f_0(980)$ mesons, respectively. However, when analyzing the Gegenbauer moments, we do not distinguish the impacts of different quark components. We estimate the Gegenbauer moments of the $f_0(500)$ meson, which are $B_1=-0.42\pm0.07$ and $B_3=0.01\pm0.04$ with the QCD sum rule in Ref. \cite{Cheng:2005nb}, respectively. Combining the total branching fraction $(15.2\pm1.4)\times10^{-6}$ in $B^-\rightarrow \pi^-\pi^+\pi^-$ measured by BaBar \cite{BaBar:2009vfr} and the fit fraction $(25.2\pm0.5\pm5.0)\%$ of the $f_0(500)$ meson \cite{LHCb:2019sus,LHCb:2019jta}, then the experimental result of decay branching fraction of $B^-\rightarrow f_0(500) \pi^-\rightarrow \pi^-\pi^+\pi^-$ decay is $(3.83\pm0.76)\times10^{-6}$ \cite{Cheng:2020hyj}. Because of the broad width of the $f_0(500)$ meson, it is not appropriate to deal with the branching fraction of the $B^-\rightarrow \pi^- f_0(500)$ decay using $\mathcal{B}(B^-\rightarrow \pi^-f_0(500) \rightarrow \pi^-\pi^+\pi^-)=\mathcal{B}(B^-\rightarrow \pi^-f_0(500))\mathcal{B}(f_0(500) \rightarrow \pi^+\pi^-)$. One can absorb the finite width effect into a factor $\xi$, that is, $\mathcal{B}(B^-\rightarrow \pi^-f_0(500) \rightarrow \pi^-\pi^+\pi^-)=\xi\mathcal{B}(B^-\rightarrow \pi^-f_0(500))\mathcal{B}(f_0(500) \rightarrow \pi^+\pi^-)$. It can be determined that $\xi=1.85$ using the method as in Ref. \cite{Qi:2018lxy}. With the $\mathcal{B}(f_0(500)\rightarrow \pi^+\pi^-)=2/3$, we obtain $\mathcal{B}(B^-\rightarrow \pi^-f_0(500)=(3.11\pm0.62)\times10^{-6}$,  which is listed in the first row and fifth column of Table \ref{branching fraction}. It is noted that the decay branching ratios for the decay channel $B^-\rightarrow \pi^-f_0(980)$ in this work differ from those in Ref. [29]. This is mainly because in Ref. [29], $|\rho_k|=0$ and $\phi_k=0$ were taken while in our work, the parameters $|\rho_k^{SP/PS}|$ and $\phi_k^{SP/PS}$ are determined by the experimental data and take different values from those in Ref. [29].

The production of $a_0(980)$ in hadronic $B$ decays has not been well detected, and only certain limitations have been set as listed in Table \ref{branching fraction}. In contrast, the production of $a_0(980)$ in charm decays has been observed in various decays, including $D^0\rightarrow \pi^+\pi^-\eta$ and $D^+\rightarrow \pi^+\pi^0\eta$ by BESIII \cite{BESIII:2024tpv}. With the values of $|\rho_k^{SP}|$ and $|\rho_k^{PS}|$ fitted in present work, our predicted results all fall within the constraints of experimental limitations. We also show the branching ratios of the $B\rightarrow a_0(1450)\pi (K)$ and $B\rightarrow K_0^*(1430)\pi (K)$ decays in Table \ref{branching fraction}. We adopt the scenario that  $a_0(1450)$  and $K_0^*(1430)$ are the first excited states of low lying light $q\bar{q}$ scalars $a_0(980)$ and $K_0^*(700)$ in our calculations. The data of $B\rightarrow a_0(1450)\pi (K)$ and $B\rightarrow K_0^*(1430)\pi (K)$ can be accommodated within the framework of QCDF when considering the contribution from the power corrections due to weak-annihilation, for example, most of the predictions of  $B\rightarrow K_0^*(1430)\pi (K)$ are consistent with the experimental data except the decay $B^-\rightarrow K^- K^{*}_0(1430)^0$, which is $\mathcal{B}(B^-\rightarrow K^- K^{*}_0(1430)^0)=(4.07\pm0.62)\times10^{-6}$ and agrees with the previous works, $(3.19^{+0.18+0.36+1.73}_{-0.14-0.62-1.39})\times10^{-6}$ \cite{Chen:2021oul}, $(3.37^{+1.03+0.55+0.34}_{-0.85-0.48-0.39})\times10^{-6}$ \cite{Li:2015zra} in QCDF and $(3.99\pm1.38\pm0.43)\times10^{-6}$ in pQCD \cite{Wang:2020saq}. Because of the large ratio between the chiral factors $\gamma_\chi^{K_0^{*0}}$ and $\gamma_\chi^{K^0}$ ($\approx 9:1$) using Eq. (\ref{rp}), one can observe that the  branching ratio of the $B^-\rightarrow K^- K^{*}_0(1430)^0$ decay is much larger than that of the $B^-\rightarrow K^0 K^{*}_0(1430)^-$ in Table \ref{branching fraction}. We also calculated the $CP$ violating asymmetries of  $B\rightarrow SP$ and the results are summarized in Table \ref{Dcp}. It can be seen that our theoretical results  agree with other theoretical results or existing experimental data. This indicates that larger values of $|\rho_k|$ fitted in present work may be reasonable and acceptable.

\begin{table}[tb]
\renewcommand{\arraystretch}{1.2}
\centering
\caption{Branching fractions (in units of $10^{-6}$) of $B$ decays to final states containing scalar mesons. In order to compare theory with experiment for decays involving $f_0(980)$, $a_0(980)$ or $a_0(1450)$, we use $B(f_0(980)\rightarrow \pi^+\pi^-)=0.50^{+0.07}_{-0.09}$ and $B(a_0(980)\rightarrow\eta\pi)=0.845\pm0.017$ \cite{Cheng:2013fba}. Experimental results are taken from \cite{Workman:2022ynf} or \cite{HFLAV:2019otj}.}
\begin{tabular*}{\textwidth}{@{\extracolsep{\fill}}ccccc}
\hline
\hline
 decay channels                    & This work                           & QCDF                           & pQCD           &Experiment    \\
\hline
 $B^-\rightarrow \pi^- f_0(500)$ & $4.03\pm0.98$      &$5.15^{+1.31+0.86}_{-1.16-1.29}$\cite{Cheng:2022ysn}   &$-$          &$3.11\pm0.62$ \\
 $B^-\rightarrow K^- f_0(500)$  &$13.58\pm3.72$     &$-$                       &$-$                                            &$-$  \\
$B^-\rightarrow \pi^- f_0(980)$ &$2.07\pm0.47$     &$0.26_{-0.03-0.04-0.12}^{+0.04+0.05+0.18}$\cite{Cheng:2013fba}            &$2.5\pm1.0$ \cite{Zhang:2008sa}                      &$<3.0$ \\
$B^-\rightarrow K^- f_0(980)$ & $20.11\pm4.82$      &$16.1_{-1.8-1.8-11.0}^{+1.9+1.2+30.8}$\cite{Cheng:2013fba}           &$[16,18]$ \cite{Wang:2006ria}                        &$18.8^{+1.8}_{-2.0}$ \\
$B^-\rightarrow \pi^- a_0(980)^0$ &$4.75\pm0.63$    &$4.9_{-0.3-1.1-0.7}^{+0.4+1.2+0.5}$\cite{Cheng:2013fba}     &$2.8^{+0.0+0.0+0.0}_{-0.8-0.9-0.6}$\cite{Zhang:2010fcy}                       &$<6.9$   \\
$B^-\rightarrow K^- a_0(980)^0$ &$0.73\pm0.41$      &$0.34_{-0.06-0.13-0.07}^{+0.08+0.45+1.02}$ \cite{Cheng:2013fba}  &$3.5^{+0.4+0.4+1.0}_{-0.4-0.6-1.0}$\cite{Shen:2006ms}          &$<3.0$   \\
$B^-\rightarrow \pi^0 a_0(980)^-$ &$0.87\pm0.30$   &$0.70_{-0.16-0.08-0.14}^{+0.22+0.09+0.22}$\cite{Cheng:2013fba}  & $0.41^{+0.00+0.00+0.00}_{-0.13-0.14-0.12}$\cite{Zhang:2010fcy}    & $<1.7$    \\
$B^-\rightarrow \bar{K}^0 a_0(980)^-$ &$0.12\pm0.09$ & $0.08_{-0.07-0.08-0.02}^{+0.12+0.58+2.12}$\cite{Cheng:2013fba}  & $6.9^{+0.8+1.1+2.0}_{-0.7-1.1-1.7}$ \cite{Shen:2006ms}      & $<4.6$    \\
$B^-\rightarrow \pi^- a_0(1450)^0$ & $6.29\pm0.82$  &$5.1_{-0.4-1.1-1.3}^{+0.5+1.2+1.3}$ \cite{Cheng:2013fba}  & $6.01^{+2.91+1.08}_{-1.72-0.99}$\cite{Chai:2021pyp}    & $-$    \\
$B^-\rightarrow K^- a_0(1450)^0$ &$4.07\pm0.48$     &$2.2_{-0.5-1.0-1.9}^{+0.7+2.3+7.7}$\cite{Cheng:2013fba}   & $12.2^{+0.1+5.9}_{-0.1-4.7}$\cite{Chai:2021pyp}  & $-$  \\
$B^-\rightarrow \pi^0 a_0(1450)^-$& $5.01\pm0.50$  &$2.1_{-0.5-0.2-0.6}^{+0.7+0.2+0.8}$ \cite{Cheng:2013fba} & $1.24^{+0.53+0.08}_{-0.31-0.10}$\cite{Chai:2021pyp}   & $-$    \\
$B^-\rightarrow \bar{K}^0 a_0(1450)^-$ &$10.34\pm0.91$     &$4.2_{-1.2-2.1-4.2}^{+1.6+4.9+18.1}$\cite{Cheng:2013fba}   & $16.3^{+0.5+8.4}_{-0.3-6.5}$ \cite{Chai:2021pyp}  & $-$  \\
$B^-\rightarrow \pi^- \bar{K}^{*}_0(1430)^0$ & $46.32\pm3.93$ &$12.9_{-3.7-3.4-9.1}^{+4.6+4.1+38.5}$\cite{Cheng:2013fba}  & $47.6^{+11.3}_{-10.1}$\cite{Shen:2006ms}    & $39_{-5}^{+6}$   \\
 &   &$33.37_{-1.04-7.29-14.95}^{+1.48+4.28+18.52}$ \cite{Chen:2021oul}  & $36.6\pm11.3\pm3.9$ \cite{Wang:2020saq}     & $$   \\
$B^-\rightarrow K^- K^{*}_0(1430)^0$ & $4.07\pm0.62$  &$3.19_{-0.14-0.62-1.39}^{+0.18+0.36+1.73}$\cite{Chen:2021oul}   & $3.99\pm1.38\pm0.43$ \cite{Wang:2020saq} & $0.38\pm0.13$    \\
 &   &$3.37_{-0.85-0.48-0.39}^{+1.03+0.55+0.34}$\cite{Li:2015zra}   & $$      & $$    \\
$B^-\rightarrow \pi^0 K^{*}_0(1430)^-$ & $12.96\pm3.66$  &$7.4_{-1.9-1.8-5.0}^{+2.4+2.1+20.1}$\cite{Cheng:2013fba}   & $28.8^{+6.8}_{-6.1}$ \cite{Shen:2006ms}      & $11.9^{+2.0}_{-2.3}$   \\
 &   &$18.42_{-0.58-3.73-7.80}^{+0.82+2.18+9.52}$\cite{Chen:2021oul}   & $12.7\pm4.2\pm1.4$ \cite{Wang:2020saq}      & $$   \\
$B^-\rightarrow K^0 K^{*}_0(1430)^-$ &$0.23\pm0.06$   &$0.08_{-0.00+0.01+0.04}^{-0.00-0.02-0.03}$\cite{Chen:2021oul}   & $0.38\pm0.22\pm0.04$ \cite{Wang:2020saq}     & $-$   \\
&   &$0.11_{-0.04-0.06-0.09}^{+0.05+0.01+0.01}$\cite{Li:2015zra}   & $$     & $$   \\
\hline
\hline
\end{tabular*}\label{branching fraction}
\end{table}

So far, there is an important issue that needs to be verified, which is the rationality of the end-point divergence parameter for the $B\rightarrow VP/PV$ processes. Our target physical quantity is the direct $CP$ violation in the $B^\pm\rightarrow \rho(770)^0\pi^\pm$ decay process. There are numerous studies on the direct $CP$ violation in the $B^\pm\rightarrow \rho(770)^0\pi^\pm$ decay, as summarized in Ref. \cite{Cheng:2022ysn}. In the past four years, the LHCb experiment has conducted a more in-depth study of the $B^\pm\rightarrow \rho(770)^0\pi^\pm$ decay process. In 2020, three different methods were utilized for this decay process and it was found that the associated  violation is nearly zero, with the result being $\mathcal{A}_{CP}(B^\pm\rightarrow \rho(770)^0\pi^\pm)=(0.7\pm1.1\pm0.6\pm4.0)\%$ \cite{LHCb:2019sus,LHCb:2019jta} in the isobar model. In 2023, the LHCb experiment adopted a new method that does not rely on the full amplitude analysis \cite{LHCb:2022tuk}, yielding the corresponding result as $\mathcal{A}_{CP}(B^\pm\rightarrow \rho(770)^0\pi^\pm)=(-0.4\pm1.7)\%$. We have theoretically predicted the $CP$ violation in this decay, and the corresponding result is $(-0.69\pm0.38)\%$  which agrees with the experimental data. Our uncertainties come from the nonperturbative parameters and the power corrections due to soft glue interactions. In QCDF, there are two kinds of  $1/m_b$ power corrections  which come from the hard scattering and weak annihilation interactions, respectively. Their contributions to $\mathcal{A}_{CP}(B^\pm\rightarrow \rho(770)^0\pi^\pm)$ weaken rather than enhance each other , thus making $\mathcal{A}_{CP}(B^\pm\rightarrow \rho(770)^0\pi^\pm)$ compatible with zero. This indicates that using the two sets of divergence parameter values for $B\rightarrow VP/PV$ decay are appropriate, eliminating one of our major concerns.

\begin{table}[tb]
\renewcommand{\arraystretch}{1.2}
\centering
\caption{Direct $CP$ violating asymmetries (in \%)  of the $B$ decays involving scalar mesons. Experimental results are taken from \cite{Workman:2022ynf} or \cite{HFLAV:2019otj}, except for that in the first row.}
\begin{tabular*}{\textwidth}{@{\extracolsep{\fill}}ccccc}
\hline
\hline
 decay channels    & This work   & QCDF  & pQCD             &Experiment    \\
\hline
 $B^-\rightarrow \pi^- f_0(500)$ & $18.23\pm5.39$      &$15.10^{+0.31+8.36}_{-0.30-11.38}$\cite{Cheng:2022ysn}   &$-$          &$16.0\pm1.7\pm2.2$\cite{LHCb:2019sus,LHCb:2019jta} \\
 $B^-\rightarrow K^- f_0(500)$  &$-13.28\pm3.23$     &$-$                       &$-$                                            &$-$  \\
$B^-\rightarrow \pi^- f_0(980)$ &$25.01\pm6.21$     &$19.0^{+3.1+2.0+36.1}_{-2.6-2.4-28.4}$\cite{Cheng:2013fba}           &$-$                     &$-50\pm54$ \\
$B^-\rightarrow K^- f_0(980)$ & $-11.47\pm2.66$      &$1.4_{-0.2-0.4-1.4}^{+0.2+0.5+1.1}$\cite{Cheng:2013fba} &$-$  &$-9.5\pm^{+4.9}_{-4.2}$ \\
$B^-\rightarrow \pi^- a_0(980)^0$ &$-7.04\pm0.73$    &$-8.8_{-0.6-0.0-0.8}^{+0.6+0.0+0.9}$\cite{Cheng:2013fba}     &$-$                      &$-$   \\
$B^-\rightarrow K^- a_0(980)^0$ &$-15.19\pm4.68$      &$-13.3_{-2.7-6.1-40.1}^{+2.7+7.3+56.8}$ \cite{Cheng:2013fba}  &$-$         &$-$   \\
$B^-\rightarrow \pi^0 a_0(980)^-$ &$-47.81\pm6.53$   &$-32.8_{-6.2-2.8-21.6}^{+5.6+3.2+20.3}$\cite{Cheng:2013fba}  & $$  & $-$    \\
$B^-\rightarrow \bar{K}^0 a_0(980)^-$ &$7.14\pm1.23$ & $0.63_{-0.57-0.49-2.86}^{+0.74+54.22+2.49}$\cite{Cheng:2013fba}  & $-$       & $-$    \\
$B^-\rightarrow \pi^- a_0(1450)^0$ & $-12.07\pm2.04$  &$-13.1_{-0.9-0.0-4.3}^{+0.9+0.0+4.9}$ \cite{Cheng:2013fba}   & $-$   & $-$    \\
$B^-\rightarrow K^- a_0(1450)^0$ &$-5.20\pm0.83$     &$-1.79_{-0.82-0.06-17.75}^{+0.69+0.64+16.61}$\cite{Cheng:2013fba}  & $-$  & $-$  \\
$B^-\rightarrow \pi^0 a_0(1450)^-$& $-20.40\pm5.06$  &$-24.6_{-3.8-2.1-23.5}^{+3.4+2.4+25.3}$\cite{Cheng:2013fba} & $-$   & $-$    \\
$B^-\rightarrow \bar{K}^0 a_0(1450)^-$ &$-6.12\pm1.35$    &$0.22_{-0.03-0.03-47.94}^{+0.04+0.06+0.59}$\cite{Cheng:2013fba}   & $-$  & $-$  \\
$B^-\rightarrow \pi^- \bar{K}^{*}_0(1430)^0$ & $4.52\pm0.73$ &$1.3_{-0.1-0.0-4.8}^{+0.1+0.0+5.9}$ \cite{Cheng:2013fba}  & $-1.5$\cite{Shen:2006ms}    & $6.1\pm3.2$   \\
 &   &$0.57_{-0.02-0.07-0.13}^{+0.02+0.07+0.13}$ \cite{Chen:2021oul}  & $-1.3\pm0.5$ \cite{Wang:2020saq}     & $$   \\
$B^-\rightarrow K^- K^{*}_0(1430)^0$ & $-15.04\pm5.13$  &$-12.25_{-0.47-1.47-2.25}^{+0.48+1.42+2.23}$\cite{Chen:2021oul}   & $17.9\pm0.4\pm8.0\pm0.9$ \cite{Wang:2020saq}      & $10\pm17$    \\
 &   &$-2.06_{-0.65-0.69-6.67}^{+0.60+0.85+5.46}$\cite{Li:2015zra}   & $$      & $$    \\
$B^-\rightarrow \pi^0 K^{*}_0(1430)^-$ & $7.96\pm0.68$  &$3.0_{-0.4-0.4-7.8}^{+0.4+0.5+10.7}$\cite{Cheng:2013fba}   &$16.0$ \cite{Shen:2006ms}  & $26^{+18}_{-14}$   \\
 &   &$5.20_{-0.16-0.18-0.46}^{+0.16+0.37+0.64}$\cite{Chen:2021oul}   & $1.5\pm1.0$ \cite{Wang:2020saq}      & $$   \\
$B^-\rightarrow K^0 K^{*}_0(1430)^-$ &$16.09\pm5.92$   &$18.58_{-0.72-0.79-1.94}^{+0.71+1.79+2.67}$\cite{Chen:2021oul}   & $-18.4\pm5.8\pm2.7\pm5.4$ \cite{Wang:2020saq}     & $-$   \\
&   &$-5.27_{-0.59-0.57-2.82}^{+0.50+0.31+2.59}$\cite{Li:2015zra}   & $$     & $$   \\
\hline
\hline
\end{tabular*}\label{Dcp}
\end{table}

\begin{figure}[htbp]
\centering
\subfigure[]{
\includegraphics[width=3.3in]{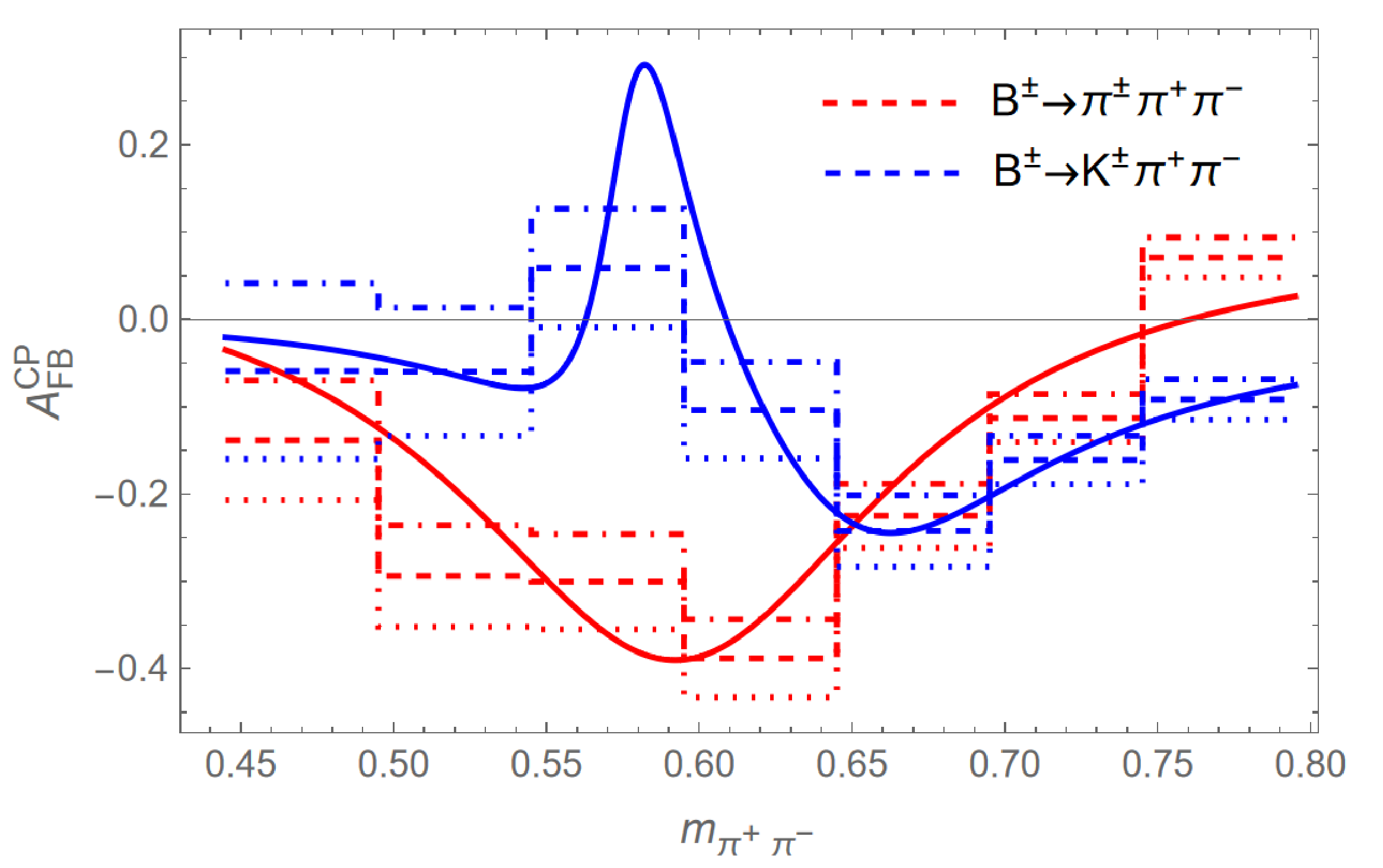}}
\subfigure[]{
\includegraphics[width=3.3in]{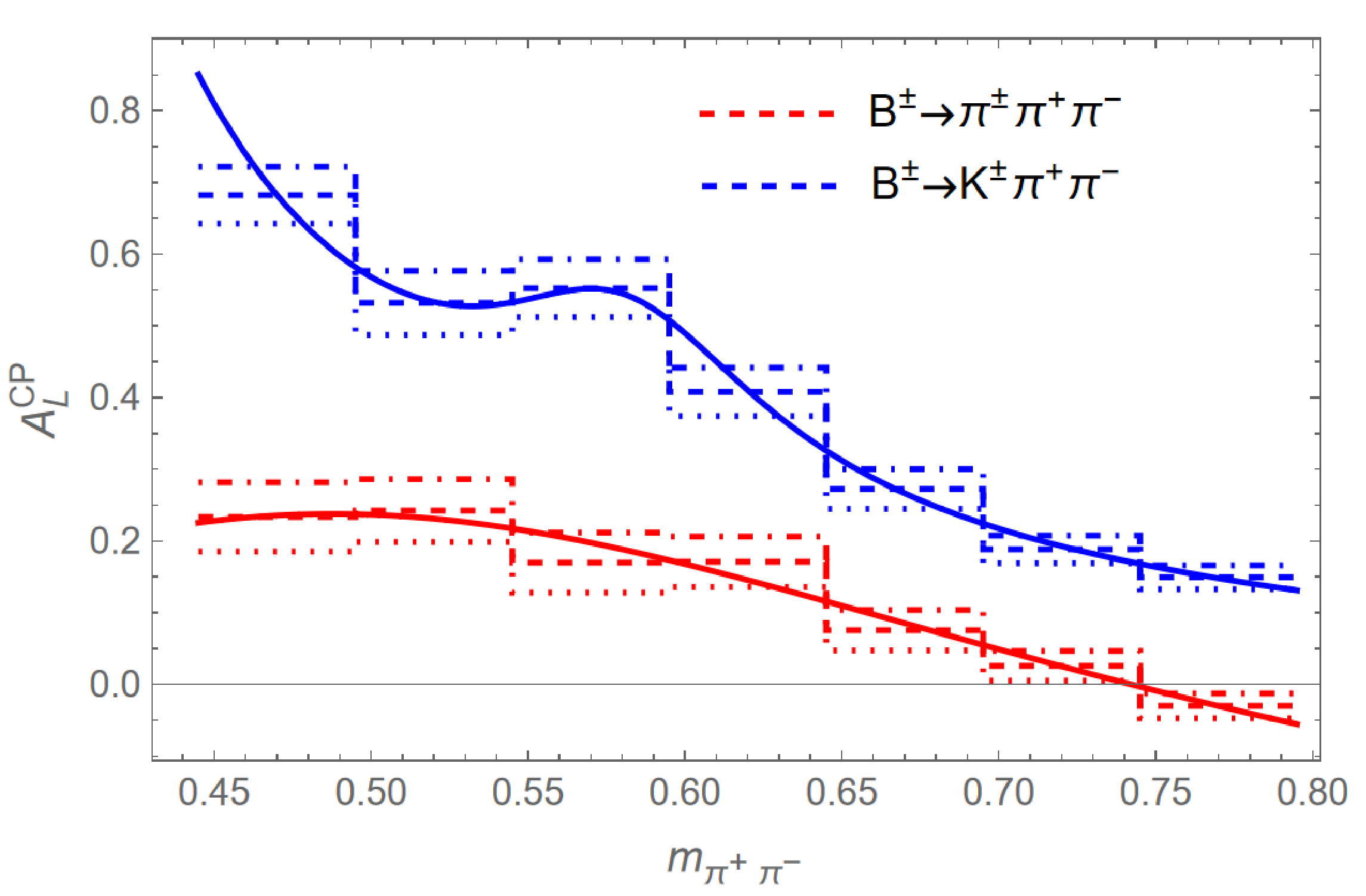}}
\caption{Comparisons of the theoretical results and experimental data for $\mathcal{A}_{FB}^{CP}$ and localized $CP$ violation for $B^\pm\rightarrow \pi^\pm\pi^+\pi^-$ and $B^\pm \rightarrow K^\pm\pi^+\pi^-$ decays, respectively, with (a) for the forward-backward asymmetry induced $CP$ asymmetries $\mathcal{A}_{FB}^{CP}$, and (b) for the localized $CP$ asymmetries $\mathcal{A}_L^{CP}$.}
\label{ACPFIG1}
\end{figure}

Based on the central values of the fitted parameters in Eq. (\ref{rhophi}), we have calculated the FBA-CPAs and the LCPAs of the $B^\pm\rightarrow \pi^\pm\pi^+\pi^-$ and $B^\pm\rightarrow K^\pm\pi^+\pi^-$ decays. Substituting Eq.(\ref{AFB2}) and the expression of its conjugate process into Eq. (\ref{AFBCP}), one can get the theoretical prediction of FBA-CPAs. Similarly, using Eq. (\ref{AFB1}), the specific form of the conjugate process, the event yields summarized in Table \ref{tabN} and Eq. (\ref{AFBCP}), the experimental results for FBA-CPAs can be obtained easily. With $\mathcal{A}_{CP}^L=\frac{\int_Lds_{12}ds_{23}(|\mathcal{M}|^2-|\bar{\mathcal{M}}|^2)}{\int_Lds_{12}ds_{23}(|\mathcal{M}|^2+|\bar{\mathcal{M}}|^2)}$, we also get the results of the LCPAs theoretically. By substituting the event yields from Table \ref{tabN} into Eq. (\ref{AFBCP}), one can derive the experimental observation of LCPAs. For comparison, both the theoretical and the experimental results for FBA-CPAs and LCPAs are presented in Fig. 3(a) and 3(b), respectively. Fig. 3 illustrates a good agreement between the theoretical and experimental results regarding the FBA-CPAs and the LCPAs,  indicating that the contribution of the $f_0(500)$ resonance is indispensable when studying asymmetrical quantities near the $\rho(770)^0$ resonance were considered.  In fact, if the interference effect from $f_0(500)$ were disregarded
, i.e., only the contribution from the $\rho(770)^0$ was considered, the theoretical results would have not matched with the experimental results \cite{Wei:2022zuf}. It is noted that FBA-CPA show peaks and troughs with maxima of 30\% and 40\% for $B^\pm\rightarrow K^\pm\pi^+\pi^-$ and $B^\pm\rightarrow \pi^\pm\pi^+\pi^-$, respectively, when the invariant mass of the $\pi\pi$ pair is around $0.60$ GeV in Fig. 3 (a). Therefore, it can be concluded that the most pronounced interference enhancements for these two decays occur at $m_{\pi\pi}\approx 0.60$ GeV for FBA-CPAs.  Within the region of our interest, LCPAs are predominantly positive for  $B^\pm\rightarrow \pi^\pm\pi^+\pi^-$ and $B^\pm\rightarrow K^\pm\pi^+\pi^-$ decays, varying from -5\% to 20\%  and 20\% to 80\%, respectively, with LCPAs relatively modest for $B^\pm\rightarrow \pi^\pm\pi^+\pi^-$. When the mass of the $\pi\pi$ pair is in the vicinity of 0.75 GeV, sign changes are observed in both FBA-CPA and LCPA for $B^\pm\rightarrow \pi^\pm\pi^+\pi^-$. However, for the $B^\pm\rightarrow K^\pm\pi^+\pi^-$ decay, only FBA-CPA undergoes a sign change while LCPA remains constant, with the invariant mass of the $\pi\pi$ pair at approximately 0.56 GeV and 0.61 GeV, respectively. In conclusion, when investigating the $CP$ asymmetry caused by the $\rho(770)^0$, the influence of the $f_0(500)$ resonance should not be disregarded.  We suggest that the above interference effect can be extended, for instance, to other decays processes of $B$ or $D$ mesons.

\section{SUMMARY}
As of now, the soft rescattering mechanism in $B$ decays involving scalar mesons has been a focus of research in hadron physics. Within the framework of the QCDF, selecting the $B^\pm\rightarrow \pi^\pm\pi^+\pi^-$ and $B^\pm\rightarrow K^\pm\pi^+\pi^-$ decay channels as the focus of this work, we have performed a global fit of the soft rescattering paremeters $|\rho_k^{SP/PS}|$ and $\phi_k^{SP/PS}$ with the aid of the event yields measured by the LHCb Collaboration. Though the fitting procedure with the interference effect between $\rho(700)^0$ and $f_0(500)$ being considered, we have obtained $|\rho_k^{SP}|=3.29\pm1.01$ and $|\rho_k^{PS}|=2.33\pm0.73$,  which are not consistent with  the traditional constraint  $|\rho_k|<1$. This seems to indicate a large soft gluon interaction effect in the $B$ decays. Using the obtained values of the $|\rho_k^{SP}|$ and $|\rho_k^{PS}|$, we have calculated the branching ratios and $CP$ violating asymmetries for some other two-body decays of the $B$ meson involving scalar mesons which are listed in Tables \ref{branching fraction} and \ref{Dcp}, respectively. The results coincide well generally with the experimental data or other theoretical  predictions. Moreover, there are experimental and theoretical studies indicating the existence of substantial annihilation contributions.  So the large values of $|\rho_k^{SP}|$ and $|\rho_k^{PS}|$ seem to be reasonable. Furthermore, we have investigated the FB-CPAs and the LACPs for the $B^\pm\rightarrow \pi^\pm\pi^+\pi^-$ and $B^\pm\rightarrow K^\pm\pi^+\pi^-$ decays. The results show evidently that the $\rho(770)^0$ and $f_0(500)$ interference mechanism can better explain the experimental data of these asymmetries in the region of $m_{\pi\pi}\in[0.445,0.795] \mathrm{GeV}$. We propose that this interference mechanism can also be extended to decays of other beauty and charmed mesons, which may help to further explore the soft gluon interactions and the structures of scalar mesons.

\acknowledgments
One of the authors (J.-J. Qi) is very grateful to Professor Hai-Yang Cheng for valuable discussions. This work was supported by  National Natural Science Foundation of China (12405115, 12105149, 12475096, 12275024, U1204115, 11775024 and 11947001), Natural Science Foundation of Zhejiang Province (LQ21A050005) and Ningbo Natural Science Foundation (2021J180).

\section{APPENDIX}
\subsection{Amplitudes of the $B$ meson two-body decays}
Within the framework of QCD factorization, the decay amplitude of $B\rightarrow M_1M_2$ can be finally given as
\begin{equation}\label{Mbrhopi}
\begin{split}
\mathcal{M}(B^-\rightarrow\rho(770)^0 \pi^-)&=\langle \rho^0 \pi^-|\mathcal{H}_{eff}|B^-\rangle\\
&=\sum_{p=u,c}\lambda_p^{(d)}\frac{G_F}{\sqrt{2}}\bigg\{\left(\alpha_1\delta_{pu}+\alpha_4^p+\alpha_{4,EW}^p\right)_{\rho^0 \pi} m_{\rho^0} A_0^{B\rightarrow\rho^0}(0)f_\pi\\
&+\left(\alpha_2\delta_{pu}-\alpha_4+\frac{3}{2}\alpha_{3,EW}^p+\frac{1}{2}\alpha_{4,EW}^p\right)_{\pi\rho^0} m_{\rho^0} F^{B\rightarrow \pi}(0)f_{\rho^0}\\
&-\left(b_2\delta_{pu}-b_3^p+\frac{1}{2}b_{3,EW}^p\right)_{\pi\rho^0} f_Bf_{\rho^0} f_\pi m_{\rho^0}/(m_Bp_c)\\
&+\left(b_2\delta_{pu}+b_3^p+b_{3,EW}^p\right)_{\rho^0\pi} f_Bf_{\rho^0} f_\pi m_{\rho^0}/(m_Bp_c) \bigg\},\\
\end{split}
\end{equation}

\begin{equation}
\begin{split}
\mathcal{M}(B^-\rightarrow\rho(770)^0 K^-)&=\langle \rho^0 K^-|\mathcal{H}_{eff}|B^-\rangle\\
&=\sum_{p=u,c}\lambda_p^{(s)}\frac{G_F}{\sqrt{2}}\bigg\{\left(\alpha_1\delta_{pu}+\alpha_4^p+\alpha_{4,EW}^p\right)_{\rho^0 K} m_{\rho^0} A_0^{B\rightarrow\rho^0}(0)f_K
+\left(\alpha_2\delta_{pu}+\frac{3}{2}\alpha_{3,EW}^p\right)_{K\rho^0}\\
&\times m_{\rho^0} F^{B\rightarrow K}(0)f_{\rho^0}+\left(b_2\delta_{pu}+b_3^p+b_{3,EW}^p\right)_{\rho^0 K} f_Bf_{\rho^0} f_Km_{\rho^0}/(m_Bp_c)\bigg\},\\
\end{split}
\end{equation}

\begin{equation}\label{Mbfpi}
\begin{split}
\mathcal{M}(B^-\rightarrow f_0(500)\pi^-)&=\langle f_0 \pi^-|\mathcal{H}_{eff}|B^-\rangle\\
&=\sum_{p=u,c}\lambda_p^{(d)}\frac{G_F}{\sqrt{2}}\bigg\{\left(\alpha_1\delta_{pu}+\alpha_4^p+
\alpha_{4,EW}^p\right)_{f_0^n\pi}(m_{f}^2-m_B^2)F_0^{B\rightarrow \pi}(0)f_\pi\\
&+\left(\alpha_2\delta_{pu}+2\alpha_{3}+\alpha_{4}+\frac{1}{2}\alpha_{3,EW}^p-\frac{1}{2}\alpha_{4,EW}^p\right)_{\pi f_0^n}(m_B^2-m_{\pi}^2)F_0^{B\rightarrow \pi}(0)\bar{f}^n_{f_0}\\
&+\left(\sqrt{2}\alpha_3^p-\sqrt{2}\alpha_{3,EW}^p\right)_{\pi f_0^s}(m_B^2-m_\pi^2)F_0^{B\rightarrow \pi}(0)\bar{f}^s_{f_0}-\left(b_2\delta_{pu}+b_3^p+b_{3,EW}^p\right)_{f_0^n \pi} f_Bf_\pi\bar{f}^n_{f_0}\\
&+\left(b_2\delta_{pu}+b_3^p-\frac{1}{2}b_{3,EW}^p\right)_{\pi f_0^n}f_Bf_\pi\bar{f}^n_{f_0}\bigg\},\\
\end{split}
\end{equation}

\begin{equation}\label{Mbfk}
\begin{split}
\mathcal{M}(B^-\rightarrow f_0(500) K^-)&=\langle f_0 K^-|\mathcal{H}_{eff}|B^-\rangle\\
&=\sum_{p=u,c}\lambda_p^{(s)}\frac{G_F}{\sqrt{2}}\bigg\{\left(\alpha_1\delta_{pu}+\alpha_4^p+
\alpha_{4,EW}^p\right)_{f_0^nK}(m_{f_0}^2-m_B^2)F_0^{B\rightarrow f_0}(0)f_K\\
&+\left(\alpha_2\delta_{pu}+2\alpha_{3}+\frac{1}{2}\alpha_{3,EW}^p)\right)_{Kf_0^n} (m_B^2-m_{K}^2)F_0^{B\rightarrow K}(0)\bar{f}^n_{f_0}\\
&+\left(\sqrt{2}\alpha_3^p+\sqrt{2}\alpha_{4}^p-\frac{1}{\sqrt{2}}\alpha_{3,EW}^p-\frac{1}{\sqrt{2}}\alpha_{4,EW}^p\right)_{Kf_0^s}(m_B^2-m_{K}^2)F_0^{B\rightarrow K}(0)\bar{f}^s_{f_0}
\\&-\left(b_2\delta_{pu}+b_3^p+b_{3,EW}^p\right)_{f_0^nK}f_Bf_K\bar{f}^n_{f_0}-\sqrt{2}\left(b_2\delta_{pu}+b_3^p+b_{3,EW}^p\right)_{Kf_0^s} f_Bf_K\bar{f}^s_{f_0}\bigg\},\\
\end{split}
\end{equation}

\begin{equation}\label{Mbfk1}
\begin{split}
\mathcal{M}(B^-\rightarrow f_0(980) \pi^-)&=\sum_{p=u,c}\lambda_p^{(d)}\frac{G_F}{\sqrt{2}}\bigg\{\left(\alpha_1\delta_{pu}+\alpha_4^p+\alpha_{4,EW}^p\right)_{ f_0^n\pi}(m_{f_0}^2-m_B^2)F_0^{B\rightarrow f_0}(0)f_\pi\\
&+\left(\alpha_3^p-\frac{1}{2}\alpha_{3,EW}^p\right)_{\pi f_0^s}(m_B^2-m_\pi^2)F_0^{B\rightarrow \pi}(0)\bar{f}_{f_0}^s\\
&+\left(\alpha_2\delta_{pu}+2\alpha_3^p+\frac{1}{2}\alpha_{3,EW}+\alpha_4^p-\frac{1}{2}\alpha_{4,EW}^p\right)_{\pi f_0^n}(m_B^2-m_\pi^2)F_0^{B\rightarrow \pi}(0)\bar{f}_{f_0}^n\\
-&\Big[\left(b_2\delta_{pu}+b_3^p+b_{3,EW}^p\right)_{f_0^n\pi}+\left(b_2\delta_{pu}+b_3^p+b_{3,EW}^p\right)_{\pi{f_0^n}}\Big]f_Bf_\pi\bar{f}_{f_0}^n\bigg\},\\
\end{split}
\end{equation}

\begin{equation}\label{Mbfk2}
\begin{split}
\mathcal{M}(B^-\rightarrow f_0(980) K^-)&=\sum_{p=u,c}\lambda_p^{(s)}\frac{G_F}{\sqrt{2}}\bigg\{\left(\alpha_3^p+\alpha_4^p-\frac{1}{2}\alpha_{3,EW}^p-\frac{1}{2}
\alpha_{4,EW}^p\right)_{K f_0^s}(m_B^2-m_K^2)F_0^{B\rightarrow K}(0)\bar{f}_{f_0}^s\\
&+\left(\alpha_1\delta_{pu}+\alpha_4^p+\alpha_{4,EW}^p\right)_{f_0^nK}(m_{f_0}^2-m_B^2)F_0^{B\rightarrow f_0}(0)f_K+\left(\alpha_2\delta_{pu}+2\alpha_3^p+\frac{1}{2}\alpha_{3,EW}\right)_{K f_0^n}\\
&\times(m_B^2-m_K^2)F_0^{B\rightarrow K}(0)\bar{f}_{f_0}^n
+\left(b_2\delta_{pu}+b_3^p+b_{3,EW}^p\right)_{f_0^nK}f_Bf_K\bar{f}_{f_0}^n\\
&+\left(b_2\delta_{pu}+b_3^p+b_{3,EW}^p\right)_{K{f_0^s}}f_Bf_K\bar{f}_{f_0}^n\bigg\},\\
\end{split}
\end{equation}
where for ease of description, we do not distinguish the decay constants,  $\bar{f}^n_{f_0}$ and $\bar{f}^s_{f_0}$, for the scalar mesons $f_0(500)$ and $f_0(980)$ in the $B^-\rightarrow f_0(500) P$ and $B^-\rightarrow f_0(980) P$ decays. However, different values for these decay constants  are assigned to $f_0(500)$ and $f_0(980)$ as shown in Table \ref{dcmo} in our calculations.

\begin{equation}\label{Mbapi}
\begin{split}
\mathcal{M}(B^-\rightarrow a_0 \pi^-)&=\sum_{p=u,c}\lambda_p^{(d)}\frac{G_F}{2}\bigg\{\left(\alpha_1\delta_{pu}+\alpha_4^p+\alpha_{4,EW}^p\right)_{ a_0\pi}(m_{a_0}^2-m_B^2)F_0^{B\rightarrow a_0}(0)f_\pi\\
&+\left(\alpha_2\delta_{pu}-\alpha_4^p+\frac{3}{2}\alpha_{3,EW}+\frac{1}{2}\alpha_{4,EW}^p\right)_{\pi a_0}(m_B^2-m_\pi^2)F_0^{B\rightarrow \pi}(0)\bar{f}_{a_0}\\
+&\Big[\left(b_2\delta_{pu}+b_3^p+b_{3,EW}^p\right)_{a_0\pi}-\left(b_2\delta_{pu}+b_3^p+b_{3,EW}^p\right)_{\pi a_0}\Big]f_Bf_\pi\bar{f}_{a_0}\bigg\},\\
\end{split}
\end{equation}

\begin{equation}\label{MbaK}
\begin{split}
\mathcal{M}(B^-\rightarrow a_0 K^-)&=\sum_{p=u,c}\lambda_p^{(s)}\frac{G_F}{2}\bigg\{\left(\alpha_1\delta_{pu}+\alpha_4^p+\alpha_{4,EW}^p\right)_{ a_0K}(m_{a_0}^2-m_B^2)F_0^{B\rightarrow a_0}(0)f_K\\
&+(a_2)_{Ka_0}(m_B^2-m_K^2)F_0^{B\rightarrow K}(0)\bar{f}_{a_0}+\left(b_2\delta_{pu}+b_3^p+b_{3,EW}^p\right)_{a_0K}f_Bf_K\bar{f}_{a_0}\bigg\},\\
\end{split}
\end{equation}
where $a_0$ stands for $a_0(980)$ or $a_0(1450)$.

\begin{equation}\label{MKpi1}
\begin{split}
\mathcal{M}(B^-\rightarrow \pi^-\bar{K}^*_0 (1430)^0 )&=\sum_{p=u,c}\lambda_p^{(s)}\frac{G_F}{\sqrt{2}}\bigg\{\left(\alpha_4^p-\frac{1}{2}\alpha_{4,EW}\right)_{ \pi K^{*}_0}(m_B^2-m_\pi^2)F_0^{B\rightarrow \pi}(0)\bar{f}_{\bar{K}^{*}_0}\\
&+\left(b_2\delta_{pu}+b_3^p+b_{3,EW}^p\right)_{ \pi \bar{K}^{*}_0}f_Bf_\pi\bar{f}_{\bar{K}^{*}_0}\bigg\},\\
\end{split}
\end{equation}

\begin{equation}\label{MKpi2}
\begin{split}
\mathcal{M}(B^-\rightarrow \pi^0K^*_0(1430)^- )&=\sum_{p=u,c}\lambda_p^{(s)}\frac{G_F}{2}\bigg\{\left(\alpha_1\delta_{pu}+\alpha_4^p+\alpha_{4,EW}^p\right)_{ \pi K^{*}_0}(m_B^2-m_K^2)F_0^{B\rightarrow \pi}(0)\bar{f}_{K^{*0}_0}\\
&+\left(b_2\delta_{pu}+b_3^p+b_{3,EW}^p\right)_{\pi K^{*}_0}f_Bf_\pi\bar{f}_{K^{*}_0}+\left(\alpha_2\delta_{pu}+\frac{3}{2}\alpha_{3,EW}^p\right)_{K^{*}_0\pi}f_Bf_\pi\bar{f}_{K^{*}_0}\bigg\},\\
\end{split}
\end{equation}

\begin{equation}\label{MKK1}
\begin{split}
\mathcal{M}(B^-\rightarrow K^-K^*_0(1430)^0  )&=\sum_{p=u,c}\lambda_p^{(d)}\frac{G_F}{\sqrt{2}}\bigg\{\left(\alpha_4^p-\frac{1}{2}\alpha_{4,EW}\right)_{ KK^{*}_0}(m_B^2-m_K^2)F_0^{B\rightarrow K}(0)\bar{f}_{K^{*}_0}\\
&+\left(b_2\delta_{pu}+b_3^p+b_{3,EW}^p\right)_{ KK^{*}_0}f_Bf_K\bar{f}_{K^{*}_0}\bigg\},\\
\end{split}
\end{equation}

\begin{equation}\label{MKK2}
\begin{split}
\mathcal{M}(B^-\rightarrow K^0K^*_0(1430)^-  )&=\sum_{p=u,c}\lambda_p^{(d)}\frac{G_F}{\sqrt{2}}\bigg\{\left(\alpha_4^p-\frac{1}{2}\alpha_{4,EW}\right)_{K^{*}_0 K}(m_B^2-m_{K^{*}_0}^2)F_0^{B\rightarrow {K^{*}_0}(0)f_K}\\
&+\left(b_2\delta_{pu}+b_3^p+b_{3,EW}^p\right)_{K^{*}_0 K}f_Bf_K\bar{f}_{K^{*}_0}\bigg\},\\
\end{split}
\end{equation}

\subsection{The propagators of the intermediate resonances}
The Gounaris-Sakurai (GS) model is widely used for handling the $\rho(770)^0$ resonance, as it enables the introduction of an analytic dispersive term to maintain unitarity at a distance from the pole mass. This model was employed by both $BABAR$ and LHCb Collaborations \cite{BaBar:2009vfr,LHCb:2019sus,LHCb:2019jta} in their analyses of the $\rho(770)^0$ resonance in the $B^\pm\rightarrow \pi^\pm\pi^+\pi^-$ decay. The GS line shape for $\rho(770)^0$  takes the form \cite{Gounaris:1968mw}
\begin{equation}
R_\rho(s)=\frac{1+D\Gamma^0_\rho/m_\rho}{(m_\rho^2-s)+f(s)-im_\rho\Gamma_\rho(s)},
\end{equation}
where
\begin{equation}
\Gamma_\rho(s)=\Gamma^0_\rho \bigg(\frac{q}{q_0}\bigg)^3\frac{m_\rho}{\sqrt{s}}X^2(|\vec{q}|r^R_{BW})\quad \mathrm{with} \quad \Gamma^0_\rho=\Gamma_\rho(m_\rho^2),
\end{equation}
and an additional mass dependence
\begin{equation}
f(s)=\Gamma_\rho^0\frac{m_\rho^2}{q_0^3}\Bigg[q^2[h(s)-h(m_\rho)]+(m_\rho^2-s)q_0^2\frac{dh}{ds}\bigg| _{m_\rho}\Bigg],
\end{equation}
where
\begin{equation}
h(s)=\frac{2}{\pi}\frac{q}{\sqrt{s}}\log\bigg(\frac{\sqrt{s}+2q}{2m_\pi}\bigg),
\end{equation}
and
\begin{equation}
\frac{dh}{ds}\bigg|_ {m_\rho}=h(m_\rho)\Bigg[(8q_0^2)^{-1}-(2m_\rho^2)^{-1}\Bigg]+(2\pi m_\rho^2)^{-1}.
\end{equation}
The constant parameter $D$ is given by
\begin{equation}
D=\frac{3}{\pi}\frac{m_\pi^2}{q_0^2}\log(\frac{m_\rho+2q_0}{2m_\pi})+\frac{m_\rho}{2\pi q_0}-\frac{m_{\pi}^2m_\rho}{\pi q_0^3}.
\end{equation}

As the $CP$ asymmetry in the $S$-wave below the inelastic ($K\bar{K}$ ) threshold seen in the Dalitz plot cannot be explained via $\pi\pi\rightarrow K\bar{K}$ rescattering, it has a different origin. The $f_0(500)$ meson is represented as a simple pole description, parametrized as \cite{Pelaez:2015qba,Oller:2004xm,LHCb:2019sus}
\begin{equation}
R_{f_0}(s)=\frac{1}{s-s_{f_0}}=\frac{1}{s-(m_{f_0}^2-\Gamma_{f_0}^2(s)/4-im_{f_0}\Gamma_{f_0}(s))},
\end{equation}
with
\begin{equation}
\Gamma_{f_0}(s)=\Gamma_{f_0}^0(\frac{q}{q_0})\frac{m_{f_0}}{\sqrt{s}}.
\end{equation}

\subsection{FLATT$\acute{\mathrm{E}}$ MODEL}
As suggested by D. V. Bugg \cite{Bugg:2008ig}, the Flatt$\acute{\mathrm{e}}$ model \cite{Flatte:1976xv} for $f_0(980)$ is slightly modified and is parametrized as
\begin{equation}\label{rho}
T_R(m_{\pi\pi})=\frac{1}{m_R^2-s_{\pi\pi}-im_R(g_{\pi\pi}\rho_{\pi\pi}+g_{KK}F_{KK}^2\rho_{KK})},
\end{equation}
where $m_R$ is the $f_0(980)$ pole mass, the parameters $g_{\pi\pi}$ and $g_{KK}$ are the $f_0(980)$ coupling constants with respect to the $\pi^+\pi^-$ and $K^+K^-$ final states, respectively, and the phase-space factors $\rho$ are given as the Lorentz-invariant phase spaces

\begin{equation}\label{f01}
\begin{split}
\rho_{\pi\pi}&=\frac{2}{3}\sqrt{1-\frac{4m_{\pi^\pm}^2}{s_{\pi\pi}}}+\frac{1}{3}\sqrt{1-\frac{4m_{\pi^0}^2}{s_{\pi\pi}}},\\
\rho_{KK}&=\frac{1}{2}\sqrt{1-\frac{4m_{K^\pm}^2}{s_{\pi\pi}}}+\frac{1}{2}\sqrt{1-\frac{4m_{K^0}^2}{s_{\pi\pi}}}.\\
 \end{split}
 \end{equation}
In Eq. (\ref{rho}), compared with the normal Flatt$\acute{\mathrm{e}}$ function, a form factor $F_{KK}=\mathrm{exp}(-\alpha k^2)$ is introduced above the $KK$ threshold and serves to reduce the $\rho_{KK}$ factor as $s_{\pi\pi}$ increases, where $k$ is momentum of each kaon in the $KK$ rest frame, and $\alpha=(2.0\pm0.25)\,\mathrm{GeV}^{-2}$ \cite{Bugg:2008ig}. This parametrization slightly decreases the $f_0(980)$ width above the $KK$ threshold. The parameter $\alpha$ is fixed to be $2.0 \,\mathrm{GeV}^{-2}$, which is not very sensitive to the fit.

\subsection{LASS MODEL}
The S-wave $K^+\pi^-$ resonance at low mass is modeled using a modified LASS lineshape \cite{Aston:1987ir,Aubert:2005ce,Aaij:2018bla}, which has been widely used in experimental analyses:
\begin{equation}\label{K1430 1}
\begin{split}
T(m_{K\pi})&=\frac{m_{K\pi}}{|\vec{q}|\cot\delta_B-i|\vec{q}|}+e^{2i\delta_B}\frac{m_0\Gamma_0\frac{m_0}{|q_0|}}{m_0^2-s_{K\pi}^2-im_0\Gamma_0\frac{|\vec{q}|}{m_{K\pi}}\frac{m_0}{|q_0|}},\\
 \end{split}
 \end{equation}
with
\begin{equation}\label{K1430 2}
\begin{split}
\cot\delta_B&=\frac{1}{a|\vec{q}|}+\frac{1}{2}r|\vec{q}|,\\
 \end{split}
 \end{equation}
where the first term is an empirical term from the elastic kaon-pion scattering and the second term is the resonant contribution with a phase factor to retain unitarity. Here $m_0$ and $\Gamma_0$ are the pole mass and width of the $K_0^*(1430)$ state, respectively, $\vec{q}$ is the momentum vector of the resonance decay product measured in the resonance rest frame, and $|\vec{q_0}|$ is the value of $|\vec{q}|$ when $m_{K\pi}=m_{K_0^*(1430)}$. In Eq. (\ref{K1430 2}), the parameters $a=(3.1\pm1.0)\,\mathrm{GeV}^{-1}$ and $r=(7.0\pm2.4)\,\mathrm{GeV}^{-1}$, being the scattering length and effective range \cite{Aaij:2018bla}, respectively, which are universal in application for the description of different processes involving a kaon-pion pair.

\end{document}